\documentclass[aps,prd,superscriptaddress,nofootinbib,tighten,preprint]{revtex4}
\include{header}

\usepackage{amsfonts}
\usepackage{amssymb}
\usepackage{amsmath}
\pdfoutput=1
\usepackage{graphicx}
\usepackage{hyperref}
\usepackage{placeins}
\usepackage{gensymb}
\usepackage{xcolor}
\usepackage[normalem]{ulem}
\graphicspath{{fig/}}

\hypersetup{
  bookmarks=true,         
  unicode=false,          
  pdftoolbar=true,        
 pdfmenubar=true,        
 pdffitwindow=true,   
 pdfstartview={FitH},    
 pdfsubject={scalar NSI},   
 pdfnewwindow=true,      
 pdfcreator={RevTeX},
 colorlinks=true,     
 linkcolor=red,         
 citecolor=blue,        
 filecolor=black,     
 urlcolor=blue,           
  }

\newcommand{\comment}[1]{}
\begin{document}

	\title{Investigating the effects of Lorentz Invariance Violation on the CP--sensitivities of the Deep Underground Neutrino Experiment}

\author{Arnab Sarker}
\email{arnabs@tezu.ernet.in}
\affiliation{ Department of Physics, Tezpur University, Assam-784028, India}

\author{Abinash Medhi}
\email{amedhi@tezu.ernet.in}
\affiliation{ Department of Physics, Tezpur University, Assam-784028, India}

\author{Moon Moon Devi}
\email{devimm@tezu.ernet.in}
\affiliation{ Department of Physics, Tezpur University, Assam-784028, India}

\begin{abstract}
The neutrino oscillations offer great potential for probing new-physics effects beyond the Standard Model. Any additional effect on neutrino oscillations can help understand the nature of these non-standard effects. The violation of fundamental symmetries may appear as new-physics effects in various neutrino experiments. Lorentz symmetry is one such fundamental symmetry in nature, the violation of which implies a breakdown of space-time symmetry. The Lorentz Invariance Violation (LIV) is intrinsic in nature and its effects exist even in a vacuum. Neutrinos can be an intriguing probe for exploring such violations of Lorentz symmetry. The effect of violation of Lorentz Invariance can be explored through the impact on the neutrino oscillation probabilities. The effect of LIV is treated as a perturbation to the standard neutrino Hamiltonian considering the Standard Model Extension (SME) framework.
 
    In this work, we have probed the effects of LIV on the measurement of neutrino oscillations parameters considering Deep Underground Neutrino Experiment (DUNE) as a case study. The inclusion of LIV affects various neutrino oscillation parameters as it modifies the standard neutrino oscillation probabilities. We looked into the capability of DUNE in constraining the LIV parameters and then explored the impact of CPT-violating LIV terms on the mass-induced neutrino oscillation probabilities. We have also probed the influence of LIV parameters on the CP-measurement sensitivity at DUNE.  \\
\noindent \textbf{Keywords:} Neutrino Physics, Beyond Standard Model, LIV, CP--violation
\end{abstract}
	\maketitle

\section{Introduction}\label{sec:introduction}
Neutrinos are fundamental particles that interact with matter via weak interactions. Although in the Standard Model (SM) framework neutrinos are considered to be massless, the phenomena of neutrino oscillations involving the oscillations of the flavor states of neutrinos imply non-zero masses of neutrinos. The neutrino oscillations have been carefully investigated and validated by a large number of experiments \cite{Workman:2022ynf, Super-Kamiokande:1998kpq, SNO:2002tuh, T2K:2013ppw, NOvA:2019cyt}. The SM cannot account for the masses of neutrinos and hence the neutrino oscillations provide an excellent motivation for exploring physics beyond the Standard Model (BSM). The neutrino oscillations in the 3-flavor scenario are controlled by six parameters- three mixing angles ($\theta_{12}$, $\theta_{13}$, $\theta_{23}$), two mass squared splittings ($\Delta m^{2}_{21}$,  $\Delta m^{2}_{32}$) and one Dirac CP phase ($\delta_{CP}$). The major current challenges in determining the neutrino mixing mostly lies in finding the Dirac CP phase ($\delta_{CP}$), the sign of atmospheric mass splitting i.e. sign of $\Delta m^{2}_{32}$ and the octant of $\theta_{23}$. The current ongoing experiments are trying to measure the hint of CP-violation in the leptonic sector. The data coming from T2K experiment \cite{T2K:2019bcf} shows a preference for the maximal CP-violation ($\delta_{CP}$ $\sim$ $252^\circ$). Also, the data rules out the possibility of CP-conserving values ($\delta_{CP}$ = $0^\circ$, $\pm$ $\pi$) upto 3$\sigma$ CL. The T2K data shows a preference of normal hierarchy (NH) of neutrino mass over the inverted hierarchy (IH) at 1$\sigma$ CL. However, the data coming from NO$\nu$A experiment \cite{alex_himmel_2020_3959581} with a longer baseline shows a best fit of $\delta_{CP}$ = $148^\circ$ for NH. For resolving the tensions between the data from T2K and NO$\nu$A, more statistics may be required. A number of upcoming neutrino experiments with larger detectors equipped with advanced technology, are expected to boost the ongoing searches. These experiments aim at improving the precision measurement of the oscillation parameters as well as searches of new-physics.

The direct detection of neutrino oscillations paves the way for the investigation of physics beyond the Standard Model. The study of possible non-standard effects offers an excellent well motivated approach to explore new-physics beyond the SM \cite{Arguelles:2019xgp,Arguelles:2022xxa}. Such non-standard effects have the potential to impact the sensitivities of different experiments towards the precise measurement of neutrino oscillation parameters. As a result, the overall physics potential of the neutrino experiments may get affected. Some of the possible non-standard effects viz. Non-Standard Interactions (NSIs) \cite{PhysRevD.17.2369,Proceedings:2019qno,Miranda:2015dra,Farzan:2017xzy,Biggio:2009nt,PhysRevLett.124.111801,Masud:2016nuj,Masud:2016bvp,Singha:2021jkn, Medhi:2021wxj,Medhi:2022qmu}, Lorentz Invariance Violation (LIV) \cite{PhysRevD.69.016005,PhysRevD.98.112013,PhysRevD.99.104062,PhysRevD.99.123018,ARIAS2007401,LSND:2005oop,MINOS:2008fnv,MINOS:2010kat,IceCube:2010fyu,MiniBooNE:2011pix,Kaur:2020ggv,DoubleChooz:2012eiq,Majhi:2022fed}, neutrino decoherence \cite{PhysRevD.56.6648,Benatti:2000ph,PhysRevD.100.055023,PhysRevD.95.113005,Lisi:2000zt}, neutrino decay \cite{PhysRevD.92.073003,PICORETI201670,SNO:2018pvg,GOMES2015345,Coloma:2017zpg,Abrahao:2015rba,Choubey:2020dhw} etc. are being explored. This will also help in the exploration of any potential physics that lies outside the realm of the SM.

The Lorentz Invariance Violation (LIV) is a violation of fundamental symmetry which breaks down the underlying structure of space-time. This is one of the scenarios of BSM physics. Lorentz symmetry underpins the quantum field theory description of the SM, which is a gauge theory that describes the interaction of fundamental particles. The CPT invariance, which stands for Charge conjugation, Parity transform, and Time reversal symmetry, is the foundation of the SM. The violation of CPT invariance may lead to Lorentz Invariance Violation as shown in \cite{PhysRevLett.89.231602}. The deviation from fundamental symmetry  is a characteristic of Planck scale physics. It can serve as a window to search for any possible non-standard effects. 

In this work, we explore the consequences of a potential violation of fundamental symmetry i.e. LIV by investigating neutrino oscillation probabilities in long baseline neutrino experiments. The most commonly used framework for the study of the violation of Lorentz symmetry is the Standard Model Extension (SME) \cite{PhysRevD.58.116002} which is an extended version of the Standard Model. For phenomenological studies, it is possible to use the minimal Standard Model Extension theory, an effective field theory (EFT) \cite{Colladay:1996iz, Colladay:1998fq,Kostelecky:2003fs}. The Standard Model local gauge symmetry is still valid in SME, but at very high energies, new-physics effects can emerge as perturbations, whereas the standard physics is still valid at lower energies. The neutrinos being among the fundamental particles, have the scope for becoming a superior probe for examining any deviation from the standard physics. The Lorentz symmetry violation can be quantified by studying its effects on the neutrino oscillations probabilities. Most studies related to LIV and CPT violations uses the SME framework as it can test the underlying symmetries through phenomenological effects. The effects of violation of such symmetries can be observed in the neutrino oscillation probabilities. Hence, the phenomenon of neutrino oscillation can be used to investigate the violation of Lorentz symmetry \cite{PhysRevD.69.016005, PhysRevD.98.112013,PhysRevD.99.104062,PhysRevD.99.123018,PhysRevD.96.095018, Satunin2019}. This also opens a portal to explore non-standard physics beyond the SM. The minimal Standard Model Extension is the most commonly used formalism for investigating Lorentz violation.
Several experiments have placed restrictions on the LIV coefficients as listed here \cite{IceCube:2017qyp, T2K:2017ega, Super-Kamiokande:2014exs, DoubleChooz:2012eiq, MiniBooNE:2011pix, MINOS:2010kat, MINOS:2008fnv, LSND:2005oop}, and the SME framework has been used in numerous studies on LIV. The reference \cite{PhysRevD.80.076007} presents the effects of Lorentz and CPT violations on neutrino oscillations as perturbative effects. The LIV-induced contradiction can be avoided by modifying the dispersion relations as shown in the reference \cite{Barenboim:2018ctx}.
The neutral kaon system provides a precise bound on CPT Violation \cite{PhysRevD.98.030001} as $\left|m(K^{0})-m(\bar{K^{0}})\right|/m_{K}<0.6\times10^{-18}$. Since the Lagrangian contains mass-squared terms instead of absolute mass terms, rewriting the limit in the mass-squared terms as $\left|m^{2}(K^{0})-m^{2}(\bar{K^{0}})\right|<0.25$ eV$^{2}$. Unlike kaons, neutrinos are fundamental particles that can provide a direct measurement of mass squared splittings. These make neutrinos more effective for probing Lorentz and CPT violations. According to the reference \cite{BARENBOIM2018631}, neutrinos can also provide a tight bound on CPT Violation. The authors of the paper \cite{Tortola:2020ncu} have put bounds on CPT-invariance violation at 3$\sigma$ as $\left|\triangle m_{21}^{2}-\triangle\bar{m}_{21}^{2}\right|<4.7\times10^{-5}$ eV$^{2}$ and $\left|\triangle m_{31}^{2}-\triangle\bar{m}_{31}^{2}\right|<2.5\times10^{-4}$ eV$^{2}$. Some bounds on the Lorentz violating coefficients can be seen from the MINOS experiment as shown in the paper \cite{PhysRevLett.110.251801} as $\left|\triangle m_{31}^{2}-\triangle\bar{m}_{31}^{2}\right|<0.8\times10^{-3}$ eV$^{2}$.

A number of recent studies with long-baseline (LBL) experiments have investigated the effects of LIV on neutrino oscillation experiments. In \cite{Majhi:2019tfi}, the impact of LIV and CPT violating parameters on the appearance and disappearance probability were investigated using NOvA. This study also showed that the presence of LIV has a significant impact on the sensitivity of the experiments (NO$\nu$A \cite{NOvA:2004blv}, T2K \cite{T2K:2011qtm}) for CP-violation and mass hierarchy studies. Additionally, the synergy of NO$\nu$A and T2K has shown a significantly enhanced sensitivities. In a separate study \cite{Agarwalla2020}, the authors took into account the DUNE experiment and investigated how LIV affected the measurement of $\theta_{23}$ octant measurement and the reconstruction of the CP phase. This study demonstrated that the presence of non-zero LIV coefficients reduces octant sensitivity, although in the presence of both $a_{e\mu}$ and $a_{e\tau}$, the octant sensitivity is restored due to their mutual nullifying effect. In \cite{Rahaman:2021leu}, the authors demonstrated that the tension between T2K and NovA is reduced in presence of LIV. Conversely, the octant and mass hierarchy sensitivity of both experiments get deteriorated by LIV parameters.

In this work, we have explored the effects of LIV parameters on the physics reach of LBL experiments taking Deep Underground Neutrino Experiment (DUNE) as a case study. The whole analysis has been performed in a model-independent way. We explore the impact of Lorentz Invariance Violation which arises as a sub-dominant effect on the neutrino oscillation probabilities. In this study, we consider one non--zero LIV parameter at a time and investigate the changes in oscillation probabilities using the theoretical framework that has been discussed qualitatively in section \ref{sec:Formalism}. We probe the capability of the experiment to constrain these LIV parameters. We then study the CP-Violation sensitivity of DUNE in the presence of the off-diagonal LIV elements. The CP-precision study has also been done to observe impact of LIV towards constraining the $\delta_{CP}$ at DUNE.  We substantially discuss the significant features observed in the sensitivity analysis for the various LIV parameters. The precise determination of the mixing parameters through LBL experiments may get affected by the inclusion of new-physics scenarios and it becomes crucial to quantify as well as constrain such effects.

This manuscript has been organized in the following manner. The theoretical framework for LIV is briefly discussed in section \ref{sec:Formalism}. In section \ref{sec:Methodology}, we outline the methodology, explore the oscillation probabilities in presence of LIV and  describe the details of DUNE that we have taken for inputs to the simulation. The main results are presented in section \ref{sec:results}, where we discuss our findings qualitatively. We then summarize the work in section \ref{sec:conclusion}.

\section{Formalism}\label{sec:Formalism}
The Lorentz symmetry is a fundamental symmetry of nature which implies invariance under the Lorentz transformations. It signifies that equations that hold in one inertial frame will also hold in any other inertial frame, as laws of nature do not depend on the perspective of the observer.
Any violation of Lorentz symmetry would lead to the breakdown of the space-time symmetry. In order to probe such violations in neutrino sector, a tiny deviation from the symmetry may be incorporated as a perturbation to the standard Hamiltonian of the neutrinos using the SME framework and then the effect on the neutrino oscillation probabilities may be looked into. We explore the effects of LIV on the appearance and disappearance probability channels in the long-baseline sector, focusing at the DUNE experiment. The formalism used in this study has been described in the following.
We consider a spinor field $\psi_{i}$ with $i$ ranging over the $N$ spinor flavors. When combined with the spinor's charge conjugate $(\psi_{i}^{C}=C\bar{\psi}_{i}^{T})$, it creates a $2N$ dimensional spinor that can be written as,
\begin{equation}
\Psi_{A}=\left(\begin{array}{c}
\psi_{i}\\
\psi_{i}^{C}
\end{array}\right).
\end{equation}
\\
In SME framework, LIV is considered as a small perturbation to the standard formalism of mass-induced neutrino oscillations. The general Lagrangian density that incorporates Lorentz Invariance Violation and CPT violation \cite{PhysRevD.85.096005,PhysRevD.58.116002,Colladay:1996iz,Colladay:1998fq,Kostelecky:2003fs} can be written as,
\begin{equation}\label{eq:Lag}
\mathcal{L}=\frac{1}{2}\bar{\Psi}_{A}(\gamma^{\mu}i\partial_{\mu}\delta_{AB}-M_{AB}+\mathbb{Q}_{AB})\Psi_{B}+h.c.
\end{equation}
In equation \ref{eq:Lag}, the first two terms represent the kinetic term and the arbitrary mass matrix $M_{AB}$ term respectively. The term $\mathbb{Q}_{AB}$ represents the Lorentz violating operator that incorporates LIV in the framework. As the possible effects due to this term would be generally small, it is treated as a perturbation. In this approach, neutrino oscillations are predominantly caused by the mass matrix while the Lorentz violation term is a perturbative effect. This gives us an way to explore potential new physics beyond the Standard Model by studying the neutrino oscillations, since it is susceptible to unconventional couplings due to its interferometric nature. There exists several SME models that can be used for exploring Lorentz symmetry breaking \cite{PhysRevD.85.096005,MINOS:2008fnv,MINOS:2010kat,MiniBooNE:2011pix,DoubleChooz:2012eiq,Super-Kamiokande:2014exs,T2K:2017ega,IceCube:2017qyp,Lin:2021cst}. We focus primarily on Lorentz-violating operators of renormalizable dimensions, which dominate low-energy physics in standard theories. For renormalizable SME with non-zero LIV coefficients, the LIV terms are restricted to only those with mass dimension $\leq 4$ which is known as the minimal Standard Model Extension (SME) framework. This framework treats the effects of LIV as perturbative in nature with minimal influence \cite{PhysRevD.58.116002}. 

In this work, we investigate the neutrino behavior within the SME framework for LIV. The Lagrangian density representing only the LIV contribution \cite{PhysRevD.85.096005,Kostelecky:2003cr} can be written as,
\begin{equation}
    \mathcal{L}=-\frac{1}{2}\left[p_{\alpha\beta}^{\mu}\bar{\psi_{\alpha}}\gamma_{\mu}\psi_{\beta}+q_{\alpha\beta}^{\mu}\bar{\psi_{\alpha}}\gamma_{5}\gamma_{\mu}\psi_{\beta}-ir_{\alpha\beta}^{\mu\nu}\bar{\psi_{\alpha}}\gamma_{\mu}\partial_{\nu}\psi_{\beta}-is_{\alpha\beta}^{\mu\nu}\bar{\psi_{\alpha}}\gamma_{5}\gamma_{\mu}\partial_{\nu}\psi_{\beta}\right]+h.c,
\end{equation}
where $p_{\alpha\beta}^{\mu}$, $q_{\alpha\beta}^{\mu}$, $r_{\alpha\beta}^{\mu\nu}$ and $s_{\alpha\beta}^{\mu\nu}$ are the Lorentz symmetry breaking parameters defined in the flavor basis. Considering only the interactions of left-handed neutrinos, the above quantities can be parameterized in the form \cite{PhysRevD.85.096005},
\begin{equation}
\begin{split}
(a_{L})_{\alpha\beta}^{\mu}&=(p+q)_{\alpha\beta}^{\mu},\\
(c_{L})_{\alpha\beta}^{\mu\nu}&=(r+s)_{\alpha\beta}^{\mu\nu}
\end{split}
\end{equation}
These are hermitian matrices defined in the flavor basis and they appear in the Hamiltonian as additional terms affecting the standard neutrino oscillations. The term $a_{\alpha\beta}^{\mu}$ is the CPT-violating LIV term whereas, $c_{\alpha\beta}^{\mu}$ is the CPT-conserving LIV term. For simplicity, in this work, we restrict to the direction independent model, i.e., we consider only the time-components of the LIV parameters. A choice of reference frame is crucial for studies related to Lorentz Invariance. Therefore, we choose the conventional Sun-centered celestial equatorial frame with $Z$--axis along the Earth's axis of rotation \cite{PhysRevD.69.016005}. This is taken as the standard frame to report measurements as shown in reference \cite{PhysRevD.66.056005}. We also consider a special limit, i.e., the restriction of rotational symmetry to reduce the complexity of problem and provide a simpler framework to study the LIV effects on neutrino oscillations. We take only the temporal dependence of LIV parameters that sets $\mu=0$ and $\nu=0$. From here on, the time-component of LIV parameters $a_{\alpha\beta}^{0}$ and $c_{\alpha\beta}^{00}$ will be simply denoted by $a_{\alpha\beta}$ and $c_{\alpha\beta}$ respectively. The CPT-odd effects in the LIV Hamiltonian is governed by $a_{\alpha\beta}$, while the CPT-even effects are governed by $c_{\alpha\beta}$. The coefficients of the hermitian LIV matrix are of mass dimensions 1 and 0 respectively, and they contribute only to $\nu-\nu$ and $\bar{\nu}-\bar{\nu}$ mixing. The CPT-odd term violates LIV explicitly which introduces a preferred direction in space-time.

According to SM, the neutrinos interact with matter via Charge Current (CC) and Neutral Current (NC) weak interactions mediating a $W^{\pm}$ and a $Z^{0}$ boson respectively. The standard Hamiltonian for neutrinos propagating through matter can be written as,
\begin{equation}
\begin{split}
H_{SI} &=H_{v}+H_{m}\\
 &=\frac{1}{2E}U\left(\begin{array}{ccc}
m_{1}^{2} & 0 & 0\\
0 & m_{2}^{2} & 0\\
0 & 0 & m_{3}^{2}
\end{array}\right)U^\dagger+\sqrt{2}G_{f}N_{e}\left(\begin{array}{ccc}
1 & 0 & 0\\
0 & 0 & 0\\ 
0 & 0 & 0
\end{array}\right) 
\end{split}
\end{equation}
where,
\begin{itemize}
    \item $H_{v}$ is the vacuum Hamiltonian.
    \item $H_{m}$ is the contribution from the interaction of neutrinos with matter.
    \item U is the Pontecorvo-Maki-Nakagawa-Sakata (PMNS) mixing matrix \cite{Maki:1962mu,Pontecorvo:1957cp,Pontecorvo:1957qd,Pontecorvo:1967fh}
    \item $G_{f}$ is Fermi's constant.
    \item $N_{e}$ is the electron number density.
    \item E represents neutrino energy.
    \item $m_{i}(i=1,2,3)$ are the mass eigenstates of neutrinos.
\end{itemize}

Considering the minimal Standard Model Extension framework (SME) \cite{PhysRevD.58.116002}, the Lorentz symmetry violation can be incorporated as a perturbation to the standard Hamiltonian, $H_{SI}$. In presence of LIV, the effective Hamiltonian with Lorentz symmetry violating components can be written as,

\begin{equation}\label{eq:Hamil}
\begin{split}
H_{eff} &= H_{SI}+\left[H_{LIV}^{CPT-}+H_{LIV}^{CPT+}\right], \\
 &= H_{SI}+\left[\left(\begin{array}{ccc}
a_{ee} & a_{e\mu} & a_{e\tau}\\
a_{e\mu}^{*} & a_{\mu\mu} & a_{\mu\tau}\\
a_{e\tau}^{*} & a_{\mu\tau}^{*} & a_{\tau\tau}
\end{array}\right)-\frac{4}{3}E\left(\begin{array}{ccc}
c_{ee} & c_{e\mu} & c_{e\tau}\\
c_{e\mu}^{*} & c_{\mu\mu} & c_{\mu\tau}\\
c_{e\tau}^{*} & c_{\mu\tau}^{*} & c_{\tau\tau}
\end{array}\right)\right].
\end{split}
\end{equation}
\\

In equation \ref{eq:Hamil}, the $H_{LIV}^{CPT_{-}}$ and $H_{LIV}^{CPT_{+}}$ terms represents the CPT-violating and CPT-conserving contributions of LIV to the Hamiltonian. The CPT-violating matrix $(H_{LIV}^{CPT_{-}})$ is parameterized using $a_{\alpha\beta}$ whereas, the CPT-conserving matrix $(H_{LIV}^{CPT_{+}})$ is parameterized using $c_{\alpha\beta}$. The form of the parametrization can be seen in the equation \ref{eq:Hamil}. The parameters $a_{\alpha\beta}$ and $c_{\alpha\beta}$ will quantify the effects of Lorentz symmetry violation, where $\alpha, \beta=e,\mu,\tau$. 
The off-diagonal elements can be generally parameterized as $a_{\alpha\beta}=\left|a_{\alpha\beta}\right|e^{i\phi_{\alpha\beta}}$. The non-observability of the Minkowski trace of $c_{L}$ causes the components xx, yy, and zz to be related to the 00 component, which results in the factor $-\frac{4}{3}$ in front of the second term \cite{Kostelecky:2003cr}. We observe from the effective Hamiltonian that LIV may arise due to either CPT-odd or CPT-even terms. In this work, we address the effects for the CPT-violating Lorentz violation terms, i.e., $a_{\alpha\beta}$ parameters.

The compelling experimental evidences of neutrino oscillations have demonstrated the prospects for physics beyond the SM. Lorentz symmetry is a fundamental property of nature. It has two kinds of transformations- rotational symmetry and boost symmetry. The breaking of Lorentz symmetry implies that properties in all directions in space-time are not equivalent. Our primary goal is to examine the possibility of LIV through CPT-violating terms in the neutrino sector. The study of Lorentz symmetry  violation via neutrino oscillations can also provide a glimpse into Planck-scale physics. We investigate the effects of LIV on neutrino oscillations using the SME framework. We use eq. \ref{eq:Hamil} to explore the consequences of Lorentz symmetry-breaking terms on neutrino oscillation probabilities. 
The DUNE experiment is taken as a case study to explore the measurement sensitivity of different oscillation parameters in long baseline experiments in the presence of LIV terms.

The similarity between the CPT-Violating LIV parameters and non-standard interaction (NSI) parameters can be shown as \cite{Ohlsson:2012kf, Diaz:2015dxa},
\begin{equation}\label{eq:rel_nsi_liv}
a_{\alpha\beta}\leftrightarrow\sqrt{2}G_{f}N_{e}\varepsilon_{\alpha\beta},
\end{equation}
where $\varepsilon_{\alpha\beta}$ and $a_{\alpha\beta}$ represents the NSI and LIV parameters respectively. Although the equation \ref{eq:rel_nsi_liv} brings an equivalence between matter NSI and LIV parameters, the underlying physics controlled by the corresponding parameters remain different. It may be noted that LIV is a fundamental effect that can occur even in a vacuum and its effect is independent of the propagating medium. On the other hand, NSI arises due to neutrino-matter interactions and it requires neutrinos to propagate through matter \cite{Diaz:2015dxa}. In presence of matter, the experimental bounds on NSI parameters may help to constrain the LIV parameters and vice versa as shown in reference \cite{Diaz:2015dxa}. The capability of various neutrino experiments towards distinguishing these phenomenological signatures in neutrinos are explored by different groups \cite{Diaz:2015dxa,Barenboim:2018lpo,Majhi:2022fed,Sahoo:2022rns}.

\begin{table}[h]
    \centering
    \begin{tabular}{|c|c|c|c|}
    \hline 
    Parameters & Bound Values [95\% CL] & Parameters & Bound Values [95\% CL]\tabularnewline
    \hline 
    \hline 
    $\left|a_{e\mu}\right|$ & $<2.56\times10^{-23}$GeV \cite{PhysRevD.91.052003} & $a_{ee}$ & $[-55, 32.5]\times10^{-23}$GeV \cite{Majhi:2019tfi}\tabularnewline
    \hline 
    $\left|a_{e\tau}\right|$ & $<4.96\times10^{-23}$GeV \cite{PhysRevD.91.052003}  & $a_{\mu\mu}$ & $[-10.5, 11.6]\times10^{-23}$GeV \cite{Majhi:2019tfi}\tabularnewline
    \hline 
    $\left|a_{\mu\tau}\right|$ & $<8.26\times10^{-24}$GeV \cite{PhysRevD.91.052003}  & $a_{\tau\tau}$ & $[-10.9, 9.1]\times10^{-23}$GeV \cite{Majhi:2019tfi}\tabularnewline
    \hline 
    \end{tabular}
    \caption{The bounds on the CPT-violating LIV parameters \cite{PhysRevD.91.052003,Majhi:2019tfi}}
    \label{tab:alimit}
\end{table}
The constraints on the coefficients for Lorentz and CPT violation in the SME framework are well tabulated in the references \cite{RevModPhys.83.11,PhysRevD.91.052003,Majhi:2019tfi}. 
In table \ref{tab:alimit}, we show the typical constraints for the diagonal \cite{Majhi:2019tfi} and off-diagonal parameters \cite{PhysRevD.91.052003} at 95$\%$ CL.
\section{Methodology}\label{sec:Methodology}
 We explore the effects of LIV parameters $a_{\alpha\beta}$ and corresponding phases $\phi_{\alpha\beta}$ on the appearance and disappearance probability channels, using first-order analytical probability expressions, in \ref{sec:Probability}. We then study the impact of $a_{\alpha\beta}$ on the $\nu$-oscillation probabilities (calculated numerically) for a baseline of 1300 km in \ref{sec:delPmue}. In \ref{sec:Simulation}, we describe the technical details of the DUNE experiment using which we have further explored LIV.

\subsection{Exploring LIV through probability channels}\label{sec:Probability}
The most relevant oscillation channels for the long-baseline neutrino experiments are appearance ($\nu_\mu$ $\rightarrow$ $\nu_e$) and disappearance ($\nu_\mu$ $\rightarrow$ $\nu_\mu$) channels.  In this section, we discuss the impact of LIV parameters on the appearance ($P_{\mu e}$) and disappearance ($P_{\mu \mu}$) probabilities. The time-evolution equation of the neutrinos in presence of LIV can be framed as,

\begin{equation}
i\frac{d}{dx} \nu_{\alpha} = H_{eff} \nu_{\alpha},
\end{equation}
where, $H_{eff}$ is the effective Hamiltonian of neutrinos as shown in equation \ref{eq:Hamil}, while $\nu_{\alpha} (\alpha =e, \mu, \tau)$ are the three neutrino flavors.

The transition probability of neutrinos from the neutrino flavor $\nu_\alpha$ to $\nu_\beta$ can be written as,
\begin{equation}
P_{\alpha\beta}=\left|\left\langle \nu_{\beta}\right|e^{-iH_{eff}L}\left|\nu_{\alpha}\right\rangle \right|^{2}.
\end{equation}

It may be noted that the incorporation of LIV in the standard Hamiltonian is similar to the case of the vector NSI. An equivalence may be brought between the NSI parameters ($\epsilon_{\alpha\beta}$) and the LIV parameters ($a_{\alpha\beta}$) via equation \ref{eq:rel_nsi_liv}. We may, therefore, derive the probability expressions in presence of LIV in an analogous way to that with vector NSI. The probability expressions shown below are obtained by following the matter perturbation theory described in literature \cite{Kikuchi:2008vq,Arafune:1997hd, Liao:2016hsa,Deepthi:2017gxg}. Here, we present the expressions for appearance and disappearance channels by restricting only to the first order in matter effect coefficient \textit{a}. While considering the first order of \textit{a}, the terms depending only on the first order of $a_{e\mu}$ and $a_{e\tau}$ survives in the appearance channel $P_{\mu e}$ and the approximated expression can be written in the following form.
\begin{subequations}\label{Probmue:main}
\begin{equation}
   P(\nu_{\mu}\rightarrow \nu_{e})=P_{\nu_{\mu}\rightarrow \nu_{e}}[a=0]+P_{\nu_{\mu}\rightarrow \nu_{e}}[a_{e\tau}]+P_{\nu_{\mu}\rightarrow \nu_{e}}[a_{e\mu}] \tag{\ref{Probmue:main}}
\end{equation}
where,
\begin{itemize}
    \item The leading term representing the standard oscillation probabilities $P_{\nu_{\mu}\rightarrow \nu_{e}}[a=0]$. 
\end{itemize}
    \begin{equation}\label{Probmue:a}
   \begin{split}
    P_{\nu_{\mu}\rightarrow \nu_{e}}[a=0]&=\rm sin^{2}2\theta_{13}s_{23}^{2}sin^{2}\Delta_{31}+c_{23}^{2}sin^{2}2\theta_{13}r^{2}\Delta_{31}^{2} +4J_{r}r\Delta_{31}\left[cos\delta sin2\Delta_{31}-2sin\delta sin^{2}\Delta_{31}\right]\\
    &\rm +2sin^{2}2\theta_{13}s_{23}^{2}\left(\frac{aL}{4E}\right)\left[\frac{1}{\Delta_{31}}sin^{2}\Delta_{31}-\frac{1}{2}sin2\Delta_{31}\right]
    \end{split}
    \end{equation}

\begin{itemize}
\item The term containing the first order correction in presence of $a_{e\tau}$
\end{itemize}
\begin{equation}\label{Probmue:b}
\begin{split}
P_{\nu_{\mu}\rightarrow \nu_{e}}[a_{e\tau}]&=\rm4L \Biggl[c_{23}s_{23}^{2}s_{13}\left\{ \left|a_{e\tau}\right|cos\left(\delta+\phi_{e\tau}\right)\left(\frac{sin^{2}\Delta_{31}}{\Delta_{31}}-\frac{1}{2}sin2\Delta_{31}\right)+
\left|a_{e\tau}\right|sin\left(\delta+\phi_{e\tau}\right)sin^{2}\Delta_{31}\right\}\\ &\rm-c_{12}s_{12}c_{23}^{2}s_{23}r\left\{ \left|a_{e\tau}\right|cos\phi_{e\tau}\left(\Delta_{31}-\frac{1}{2}sin2\Delta_{31}\right)-\left|a_{e\tau}\right|sin\phi_{e\tau}sin^{2}\Delta_{31}\right\} \Biggl]
\end{split}
\end{equation}

\begin{itemize}
\item The term containing the first order correction in presence of $a_{e\mu}$
\end{itemize}
\begin{equation}\label{Probmue:c}
\begin{split}
P_{\nu_{\mu}\rightarrow \nu_{e}}[a_{e\mu}]&=\rm-4L\Biggl[s_{23}s_{13}\biggl\{ \left|a_{e\mu}\right|cos\left(\delta+\phi_{e\mu}\right)\left(s_{23}^{2}\frac{sin^{2}\Delta_{31}}{\Delta_{31}}-\frac{c_{23}^{2}}{2}sin2\Delta_{31}\right)\\&\rm +c_{23}^{2}\left|a_{e\mu}\right|sin\left(\delta+\phi_{e\mu}\right)sin^{2}\Delta_{31}\biggl\}
\rm-c_{12}s_{12}c_{23}r\biggl\{ \left|a_{e\mu}\right|cos\phi_{e\mu}\left(c_{23}^{2}\Delta_{31}+\frac{s_{23}^{2}}{2}sin2\Delta_{31}\right)\\ &\rm +s_{23}^{2}\left|a_{e\mu}\right|sin\phi_{e\mu}sin^{2}\Delta_{31}\biggl\}\Biggl]
\end{split}
\end{equation}
\end{subequations}

\begin{itemize}
\item Here, 
$$\rm r=\left(\frac{\Delta m_{21}^{2}}{\Delta m_{31}^{2}}\right),\hspace{5pt} J_{r}=c_{12}s_{12}c_{13}^{2}s_{13}c_{13}s_{23},\hspace{5pt}  \Delta_{31}=\frac{\Delta m_{31}^{2}L}{4E},\hspace{5pt}  a=2\sqrt{2}G_{f}N_{e}E$$  
\end{itemize}

We see that $P_{\mu e}$ is mostly sensitive to $a_{e\mu}$, $a_{e\tau}$ and the corresponding phases $\phi_{e\mu}$, $\phi_{e\tau}$. We explore the impact of LIV parameters on the numerically calculated probabilities in section \ref{sec:delPmue}. Similarly, the disappearance probability channel $P_{\mu\mu}$ upto the first order of matter coefficient \textit{a} can be written as \cite{Kikuchi:2008vq},
\begin{subequations}\label{Probmumu:main}
\begin{equation}
   P(\nu_{\mu}\rightarrow \nu_{\mu})=P_{\nu_{\mu}\rightarrow \nu_{\mu}}[a=0]+P_{\nu_{\mu}\rightarrow \nu_{\mu}}[(a_{\mu\mu}-a_{\tau\tau})]+P_{\nu_{\mu}\rightarrow \nu_{\mu}}[a_{\mu\tau}]
   \tag{\ref{Probmumu:main}}
\end{equation}
where,
\begin{itemize}
\item The leading term representing the standard oscillation probabilities $P_{\nu_{\mu}\rightarrow \nu_{\mu}}[a=0]$.
\end{itemize}
\begin{equation}\label{Probmumu:a}
   P_{\nu_{\mu}\rightarrow \nu_{\mu}}[a=0]=1-4c_{23}^{2}s_{23}^{2}sin^{2}\Delta_{31}+4c_{23}^{2}s_{23}^{2}c_{12}^{2}r\Delta_{31}sin2\Delta_{31}
\end{equation}

\begin{itemize}
\item The term containing the first order correction for the dependence on $(a_{\mu\mu}-a_{\tau\tau})$
\end{itemize}
\begin{equation}\label{Probmumu:b}
P_{\nu_{\mu}\rightarrow \nu_{\mu}}[(a_{\mu\mu}-a_{\tau\tau})]=\left[2Lsin2\Delta_{31}-8\frac{2E}{\Delta m_{31}^{2}}sin^{2}\Delta_{31}\right]c_{23}^{2}s_{23}^{2}\left(c_{23}^{2}-s_{23}^{2}\right)\left(a_{\mu\mu}-a_{\tau\tau}\right)
\end{equation}

\begin{itemize}
\item The term containing the first order correction in presence of $a_{\mu\tau}$
\end{itemize}
\begin{equation}\label{Probmumu:c}
    P_{\nu_{\mu}\rightarrow \nu_{\mu}}[a_{\mu\tau}]=\rm \left[\left(c_{23}^{2}-s_{23}^{2}\right)^{2}\frac{2E}{\Delta m_{31}^{2}}sin^{2}\Delta_{31}-Lc_{23}^{2}s_{23}^{2}sin2\Delta_{31}\right]8c_{23}s_{23}Re\left(a_{\mu\tau}\right)
\end{equation}
\end{subequations}

As seen in equation \ref{Probmumu:main}, $P_{\mu\mu}$ is primarily sensitive to $a_{\mu\tau}$. Interestingly, we also see that it depends on the difference of diagonal elements $a_{\mu\mu}$ and $a_{\tau\tau}$ i.e. $(a_{\mu\mu}-a_{\tau\tau})$. The other parameters arise as sub-dominant terms only in the higher-order expansion. Analogous to that shown in the references \cite{Kikuchi:2008vq,Masud:2018pig,Masud:2015xva} for the NSI case, we see that $P_{\mu e}$ is most sensitive to $a_{e\mu}$, $a_{e\tau}$ and is nominally affected by $a_{ee}$. The $P_{\mu\mu}$ oscillation channel is most sensitive to the presence of $a_{\mu\tau}$ only.

Note that, the probability expressions shown above are approximated analytical equations that contain only the first-order terms. Although we have looked into these expressions to understand the impact of various $a_{\alpha\beta}$ parameters to the leading order and sub--dominant terms, the results shown in this paper are with numerically calculated probabilities. These  probabilities are exact and  also contains the effects arising from the higher-order terms. For the sake of understanding certain effects, we have included the expressions up to the second order of matter coefficient \textit{a}, for $P_{\mu e}$ and $P_{\mu \mu}$, as appendix \ref{App:secondorder}. In subsection \ref{sec:delPmue}, we now study in detail, the effect of $a_{\alpha\beta}$ on the numerically calculated $P_{\mu e}$ and $P_{\mu \mu}$ oscillation channels.

\subsection{Probing the effects of $a_{\alpha\beta}$ on exact oscillation probabilities}\label{sec:delPmue}
In this study, we particularly probe the effect of CPT-violating LIV parameters $(a_{\alpha\beta})$ on the $\nu$-oscillation probabilities. The parameter values used for the simulation are listed in table \ref{tab:parameters}.
\begin{table}[h]
    \centering
    \begin{tabular}{|c|c|c||c|c|c|}
    \hline 
    Parameters & Values & Marginalization & Parameters & Values & Marginalization\tabularnewline
    \hline 
    \hline 
    $\theta_{12} [^{\circ}]$ & 34.51 & fixed & L$\left[km\right]$ & 1300 & fixed\tabularnewline
    \hline 
    $\theta_{13} [^{\circ}]$ & 8.44 & fixed & $\delta_{CP}$ & $-\pi/2$ & $\left[-\pi,\pi\right]$\tabularnewline
    \hline 
    $\theta_{23} [^{\circ}]$ & 47 & 39--51 & Hierarchy & Normal & fixed\tabularnewline
    \hline 
    $\triangle m_{21}^{2}\left[10^{-5}eV^{2}\right]$ & 7.56 & fixed & $\triangle m_{31}^{2}\left[10^{-3}eV^{2}\right]$ & 2.55 & 2.428--2.597\tabularnewline
    \hline 
    \end{tabular}
    \caption{Values of neutrino mixing parameters used in the simulation \cite{Esteban:2020cvm} along with the marginalization range.}
    \label{tab:parameters}
\end{table}

We explore the effects of different LIV parameters on $P_{\mu e}$ and $P_{\mu \mu}$ oscillation channels for the DUNE baseline of L = 1300 km. For the incorporation of LIV, we can modify the standard Hamiltonian accordingly as shown in equation \ref{eq:Hamil}. We calculate $P_{\mu e}$ and $P_{\mu \mu}$ by using this effective Hamiltonian in a probability calculator package. Here, we use NuOscProbExact \cite{Bustamante:2019ggq}, which is a python-based package for the numerical calculation of $\nu$-oscillation probabilities using the Ohlsson-Snellman method. It computes the oscillation probabilities for any arbitrary time-independent Hamiltonian using an expansion of the evolution operator in terms of SU(2) and SU(3) matrices. 


We consider here different choices of $a_{\alpha\beta}$ taking only one non-zero parameter at a time. For convenience and ease in representation, we introduce the notation $a_{\alpha\beta}^{\prime}$ in all the figures, where,
\begin{subequations}
\begin{equation} \label{eq:prime_diagonal}
a^{\prime}_{\alpha\beta}=\rm a_{\alpha\beta}/10^{-23} GeV\hspace{15mm} (Diagonal \hspace{1mm}elements, \alpha=\beta),
\end{equation}
\begin{equation} \label{eq:prime_offdiagonal}
a^{\prime}_{\alpha\beta}=\rm \left|a_{\alpha\beta}\right|e^{i\phi}/10^{-23} GeV\hspace{5mm} ( Off-diagonal\hspace{1mm} elements, \alpha\neq\beta)
\end{equation}
\end{subequations}

\begin{figure}[h]
    \centering
    \includegraphics[width=1\linewidth]{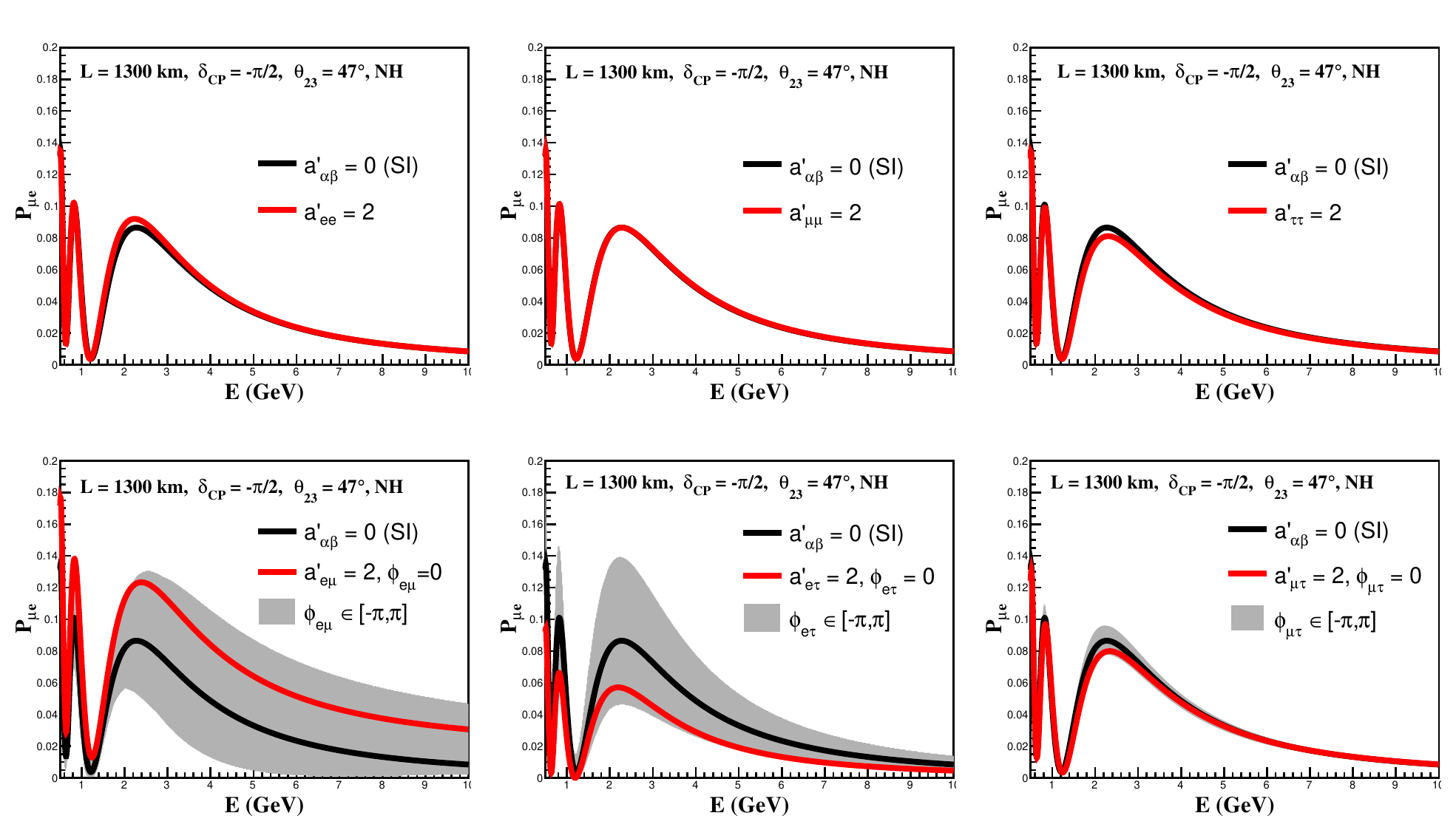}
    \caption{The impact of LIV parameters $a_{ee}$ (top-left panel), $a_{\mu\mu}$ (top-middle panel), $a_{\tau\tau}$ (top-right panel), $a_{e\mu}$ (bottom-left panel), $a_{e\tau}$ (bottom-middle panel) and $a_{\mu\tau}$ (bottom-right panel) on the $\nu_{\mu}\rightarrow\nu_{e}$ oscillation probability for DUNE with fixed $\delta_{CP}$ = -$\pi/2$ and $\theta_{23}$ = 47$^\circ$. The black solid lines represent the standard case with no LIV effects. The red solid line represents the case with $a'_{\alpha\beta}=2, \phi_{\alpha\beta}=0$ and the shaded region signifies the variation of $\phi\in[-\pi,\pi]$.}
    \label{fig:all_Pmue_band}
\end{figure}

This implies that $a_{\alpha\beta}^{\prime}$ represents the parameter values of $a_{\alpha\beta}$ expressed in units of $10^{-23}$ GeV. In figure \ref{fig:all_Pmue_band}, we perform a preliminary probe on the effects of LIV parameters on the $P_{\mu e}$ channel. Here, we consider the cases of $a_{ee}$, $a_{\mu\mu}$, $a_{\tau\tau}$, $a_{e\mu}$, $a_{e\tau}$ and $a_{\mu\tau}$ one at a time. The effects due to the presence of phase $\phi_{\alpha\beta}$ for off-diagonal elements are shown by the shaded grey band. The benchmark values of the mixing parameters used in our analysis are listed in table \ref{tab:parameters}. We present the probability values in the energy range 0.5 - 10 GeV. We consider the normal hierarchy of mass ordering with $\theta_{23}=47^{\circ}$ and $\delta_{CP}=-\pi/2$. In every sub-figure, the black solid line represents the standard no--LIV case, i.e., $a_{\alpha\beta}=0$. The red solid line represents the case with $a^{\prime}_{\alpha\beta}=2$, $\phi_{\alpha\beta}=0$.  The effects of $(a_{ee}$, $a_{\mu\mu}$, $a_{\tau\tau})$ and $(a_{e\mu}$, $a_{e\tau}$, $a_{\mu\tau})$ are shown in the top and bottom panels respectively. We observe the followings.
\begin{itemize}
    \item The presence of $a_{ee}$ $(a_{\tau\tau})$ shows a nominal enhancement (suppression) at the oscillation peak. It may also be observed from equation \ref{Probmue:main} that $P_{\mu e}$ has no dependency on $a_{ee}$ and $a_{\tau\tau}$ up to the first order. The mild dependency seen in the figure arises from the higher-order terms. For $a_{\mu\mu}$, we observe no significant changes in the probability.
    \item The presence of $a_{e\mu}$ enhances the oscillation channel $P_{\mu e}$ and shifts the oscillation peak towards higher energy. The significant effect of $a_{e\mu}$ on $P_{\mu e}$ is also validated by the first-order approximate expression shown in \ref{Probmue:c}. The shaded region shows the effect of $\phi_{e\mu}$ that can significantly modify the probability values. 
    \item For $a_{e\tau}$, we see a suppression of $P_{\mu e}$ along with a shift of the oscillation peak towards lower energy. The effects of $a_{e\tau}$ on $P_{\mu e}$ are opposite in nature to that of $a_{e\mu}$. Here, the shaded band shows the impact of phase $\phi_{e\tau}$.
    \item  The impact of $a_{\mu\tau}$ seem to nominally suppress the probability peak.  
\end{itemize}

\begin{figure}[h]
    \centering
    \includegraphics[width=1\linewidth]{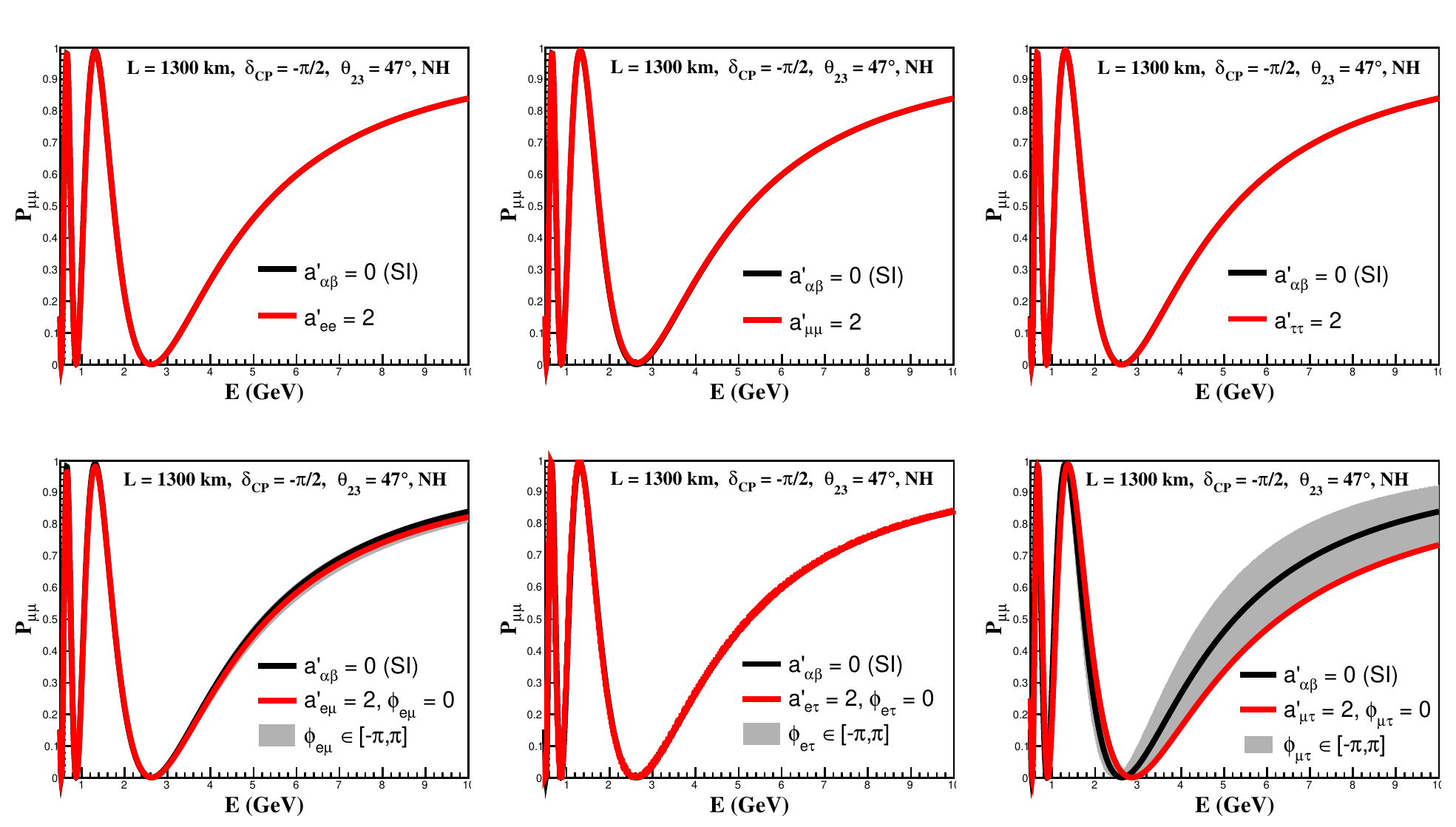}
    \caption{The impact of LIV parameters $a_{ee}$ (top-left panel), $a_{\mu\mu}$ (top-middle panel), $a_{\tau\tau}$ (top-right panel), $a_{e\mu}$ (bottom-left panel), $a_{e\tau}$ (bottom-middle panel) and $a_{\mu\tau}$ (bottom-right panel) on the $\nu_{\mu}\rightarrow\nu_{\mu}$ oscillation probability for DUNE with fixed $\delta_{CP}$ = -$\pi/2$ and $\theta_{23}$ = 47$^\circ$. The black solid lines represent the standard case with no LIV effects. The red solid line represents the case with $a'_{\alpha\beta}=2, \phi_{\alpha\beta}=0$ and the shaded region signifies the variation of $\phi\in[-\pi,\pi]$.}
    \label{fig:all_Pmumu_band}
\end{figure}

In figure \ref{fig:all_Pmumu_band}, we show the effects of LIV parameters $a_{\alpha\beta}$ on $P_{\mu\mu}$. We consider the normal hierarchy of mass ordering with $\theta_{23}=47^{\circ}$ , $\delta_{CP}=-\pi/2$ and the observations are listed below.
\begin{itemize}   
    \item The presence of $a_{\mu\tau}$ impacts $P_{\mu\mu}$, which increases with energy. We can see significant effects beyond the second oscillation minima at $\sim 2.5$ GeV. We also observe marginal energy-dependent shifts in the probability pattern, which become noticeable with an increase in the energy (e.g., see the second minima). The major contribution comes from the second term in the expression for $P_{\nu_{\mu}\rightarrow \nu_{\mu}}[a_{\mu\tau}]$ in equation \ref{Probmumu:c}.
    \item The effects from all the other parameters except $a_{\mu\tau}$ is nominal on $P_{\mu\mu}$.
\end{itemize}

\begin{figure}[t]
    \centering
    \includegraphics[width=0.49\linewidth,height=5cm]{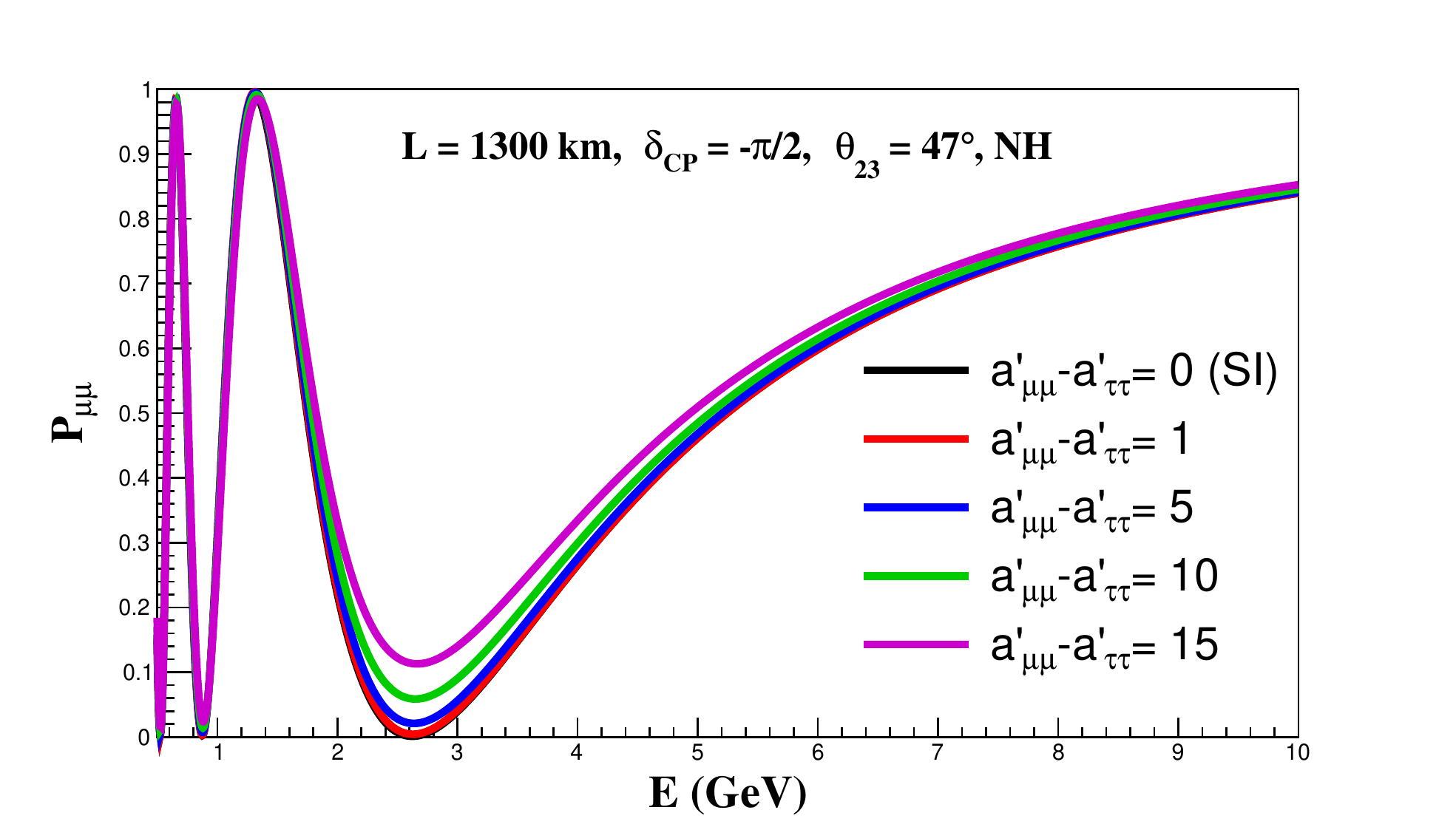}
    \includegraphics[width=0.49\linewidth,height=5cm]{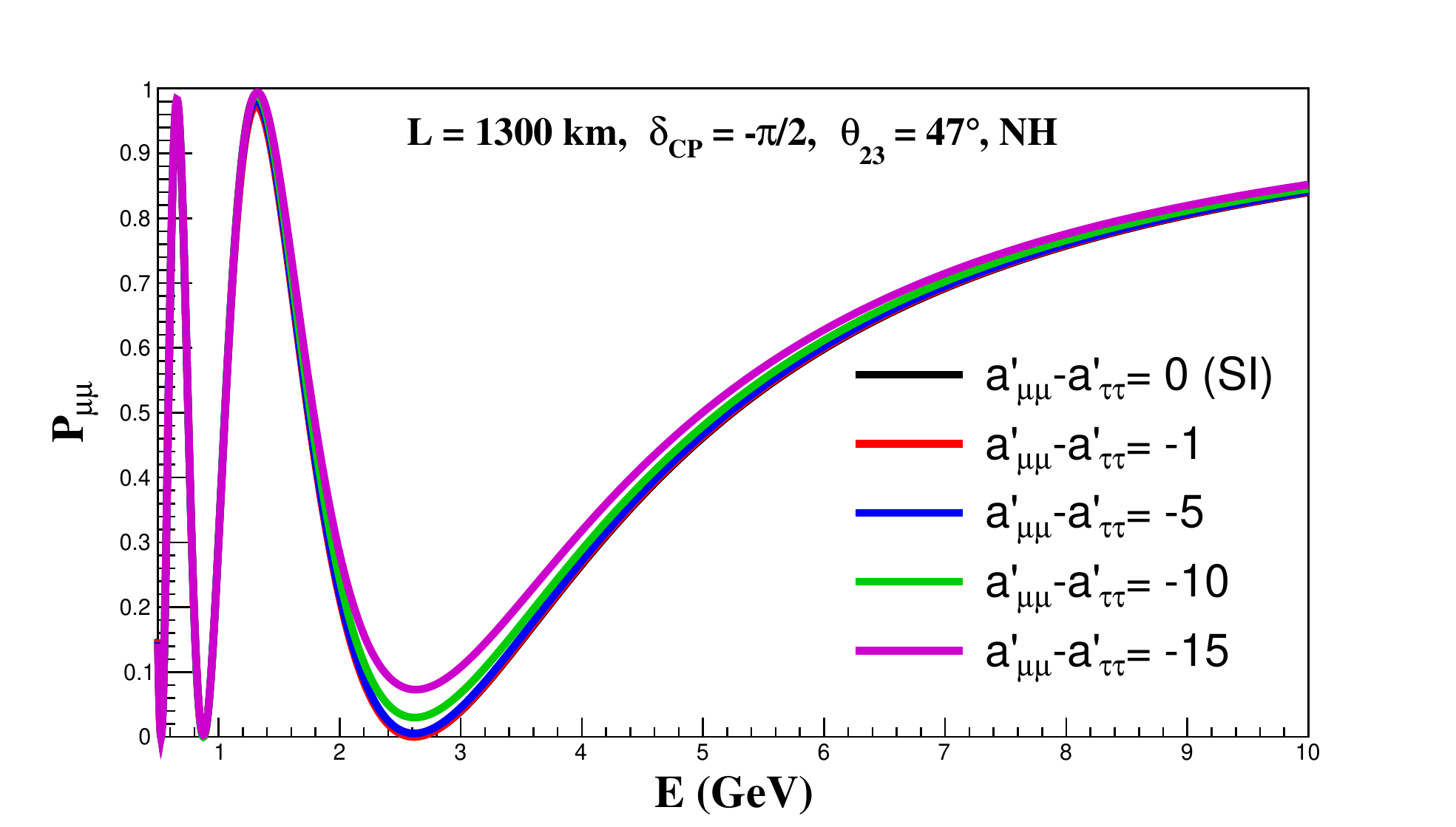}
    \caption{The impact of $(a_{\mu\mu}-a_{\tau\tau})$ on $\nu_{\mu}\rightarrow\nu_{\mu}$ oscillation probability for DUNE with fixed $\delta_{CP}$ = -$\pi/2$ and $\theta_{23}$ = 47$^\circ$ for positive (left panel) and negative (right panel) values. The black solid lines represent the standard case with no LIV effects. The other coloured lines represent the case with different values of $(a_{\mu\mu}-a_{\tau\tau})$.}
    \label{fig:Pmumu_amm_att}
\end{figure}

As in equation \ref{Probmumu:main}, we have seen a dependence of $P_{\mu\mu}$ on the difference $(a_{\mu\mu}-a_{\tau\tau})$.
In figure \ref{fig:Pmumu_amm_att}, we have studied the variation of $P_{\mu\mu}$ for non-zero value of $(a_{\mu\mu}-a_{\tau\tau})$. The different coloured lines represent different strengths of $(a_{\mu\mu}-a_{\tau\tau})$. We observe that the oscillation dip enhances as we increase the value of $(a_{\mu\mu}-a_{\tau\tau})$ for both positive (left panel) and negative (right panel) respectively.

\begin{figure}[h]
    \centering
    \includegraphics[width=1\linewidth]{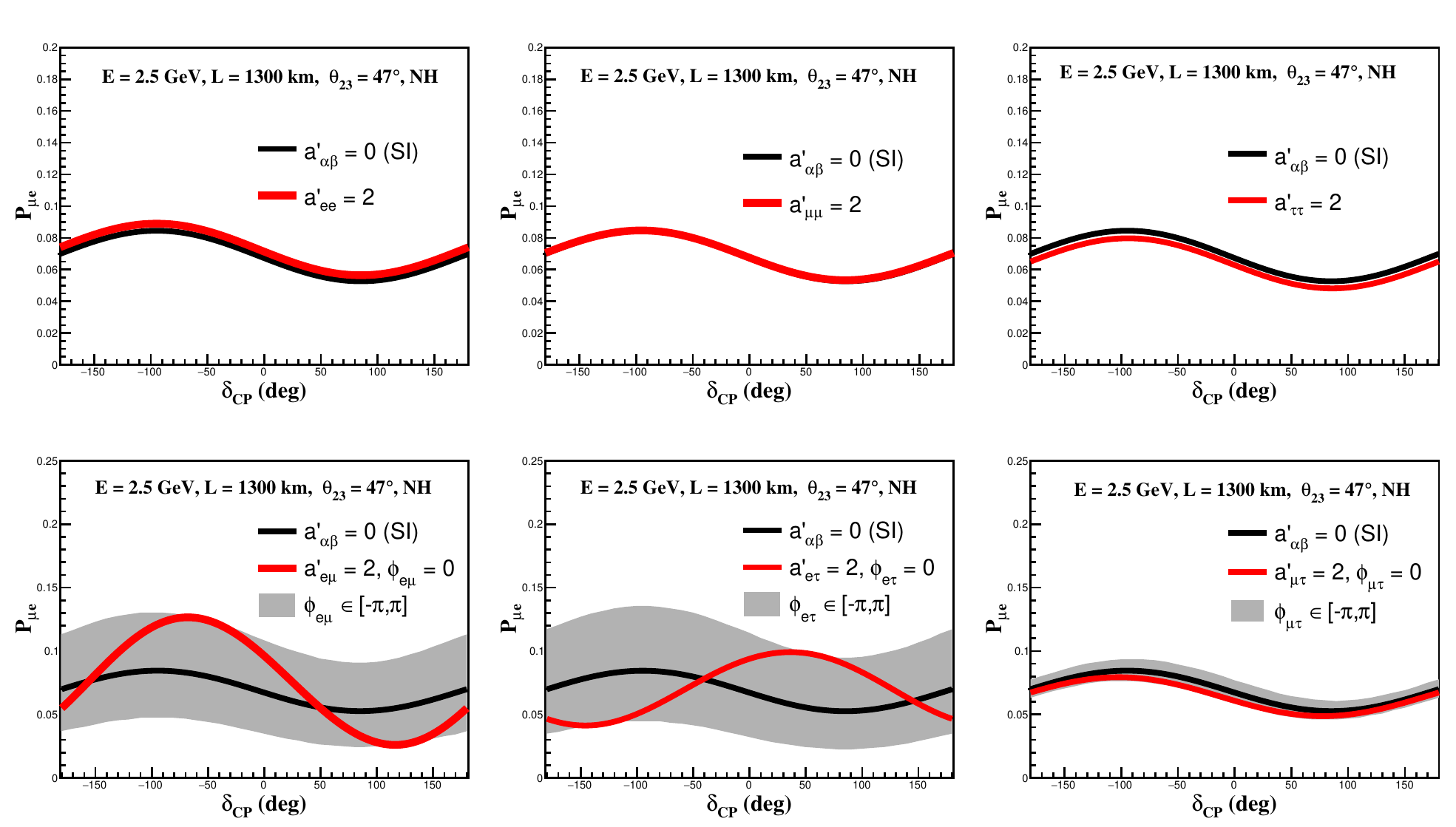}
    \caption{Plot of $P_{\mu e}$ vs $\delta_{CP}$ for diagonal elements $(a_{ee}, a_{\mu\mu}, a_{\tau\tau})$ in top-panels and off-diagonal elements $(a_{e\mu}, a_{e\tau}, a_{\mu\tau})$ in bottom panels respectively. The red solid line represents the case with  $a^{'}_{\alpha\beta}=2$, $\phi_{\alpha\beta}=0$ by fixing the energy at the first oscillation peak of DUNE i.e. E = 2.5 GeV. The black solid line represents the standard case with no LIV effect. The shaded band shows the effect of $\phi\in[-\pi,\pi]$.}
    \label{fig:Pmue_vs_dcp}
\end{figure}

\begin{figure}[h]
    \centering
    \includegraphics[width=1\linewidth]{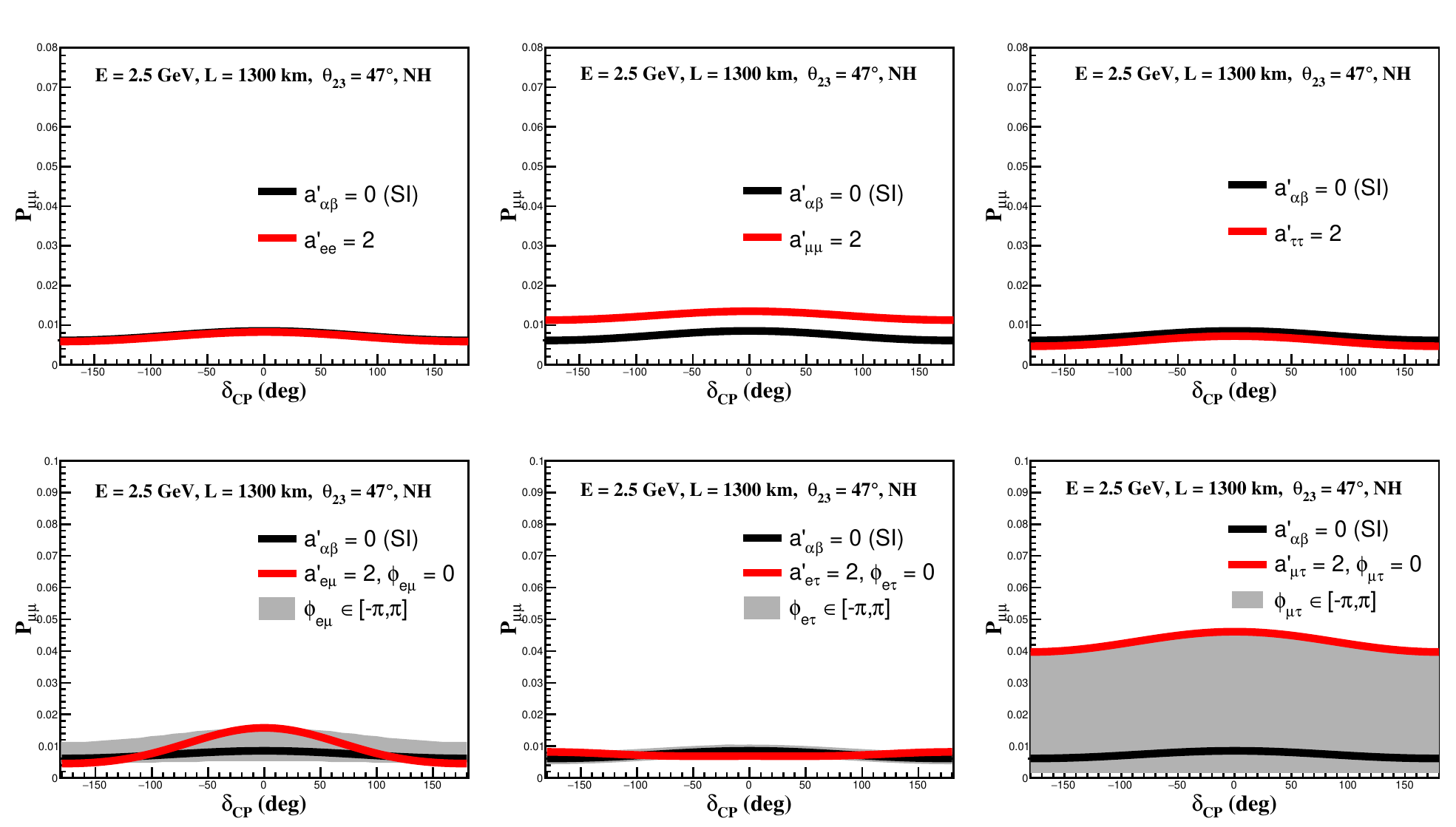}
    \caption{Plot of $P_{\mu \mu}$ vs $\delta_{CP}$ for diagonal elements $(a_{ee}, a_{\mu\mu}, a_{\tau\tau})$ in top-panels and off-diagonal elements $(a_{e\mu}, a_{e\tau}, a_{\mu\tau})$ in bottom panels. The red solid line represents the case with  $a_{\alpha\beta}=2$ and $\phi_{\alpha\beta}=0$ by fixing the energy at the first oscillation peak of DUNE i.e. E = 2.5 GeV. The black solid line represents the standard case with no LIV effect. The shaded band shows the effect of $\phi\in[-\pi,\pi]$.}
    \label{fig:Pmumu_vs_dcp}
\end{figure}

We further investigate the effects of LIV parameters on the oscillation channel $P_{\mu e}$ and $P_{\mu\mu}$ in figure \ref{fig:Pmue_vs_dcp} and \ref{fig:Pmumu_vs_dcp} respectively. We show the variation of $P_{\mu e}$ with $\delta_{CP}$ for $a_{ee}$, $a_{\mu\mu}$, $a_{\tau\tau}$ (top panels) and  $a_{e\mu}$, $a_{e\tau}$, $a_{\mu\tau}$ (bottom panels) with $\phi_{\alpha\beta}=0$, taking only one parameter at a time in figure \ref{fig:Pmue_vs_dcp}. We consider normal hierarchy for mass ordering with $\theta_{23}=47^{\circ}$ and fixing energy at the first oscillation peak of DUNE i.e. E = 2.5 GeV. The black solid line represents the case with no LIV. The red solid lines represent the case with $a_{\alpha\beta}^{\prime}=2$, $\phi_{\alpha\beta}=0$. The shaded band for off-diagonal elements is obtained by varying $\phi\in[-\pi,\pi]$. We observe the following in figure \ref{fig:Pmue_vs_dcp}.
\begin{itemize}
    \item The presence of diagonal elements $(a_{ee},a_{\mu\mu},a_{\tau\tau})$ has nominal effects on $P_{\mu e}$.
    \item In presence of $a_{e\mu}$ (bottom-left panel, figure \ref{fig:Pmue_vs_dcp}), $P_{\mu e}$ gets enhanced in the $\delta_{CP}$ range $\in \left[-150^{\circ},50^{\circ}\right]$ and gets suppressed for other $\delta_{CP}$ regions. The presence of $\phi_{e\mu}$ can bring degeneracy in $\delta_{CP}$.
    \item In presence of $a_{e\tau}$ (bottom-middle panel, figure \ref{fig:Pmue_vs_dcp}), $P_{\mu e}$ gets altered. We note that the enhancements/suppressions in $P_{\mu e}$ from its standard (no-LIV) values depend significantly on $\delta_{CP}$ -- $a_{e\tau}$ combinations.
    \item For $a_{\mu\tau}$ (bottom-right panel, figure \ref{fig:Pmue_vs_dcp}), we observe a suppression of $P_{\mu e}$ for the complete parameter space of $\delta_{CP}$. The presence of $\phi_{\mu\tau}$ nominally modifies the probability values.
\end{itemize}

In figure \ref{fig:Pmumu_vs_dcp}, we explore the effects on the disappearance channel in $\delta_{CP}\in [-\pi,\pi]$ where we observe that
\begin{itemize}
    \item The presence of diagonal elements $(a_{ee},a_{\mu\mu},a_{\tau\tau})$ has nominal effects on $P_{\mu \mu}$ with the variation of $\delta_{CP}$.
    \item In presence of $a_{\mu\tau}$ (bottom-right panel, figure \ref{fig:Pmumu_vs_dcp}), $P_{\mu \mu}$ gets enhanced in the complete $\delta_{CP}$ parameter space. However, the presence of $\phi$ suppresses the probability values.
    \item In presence of $a_{e\mu}$ (bottom-left panel) and $a_{e\tau}$ (bottom-middle panel), $P_{\mu \mu}$ gets altered nominally. A non-zero phase creates a small band near the standard probability values.
\end{itemize}

\begin{figure}[h]
    \centering
    \includegraphics[width=1\linewidth]{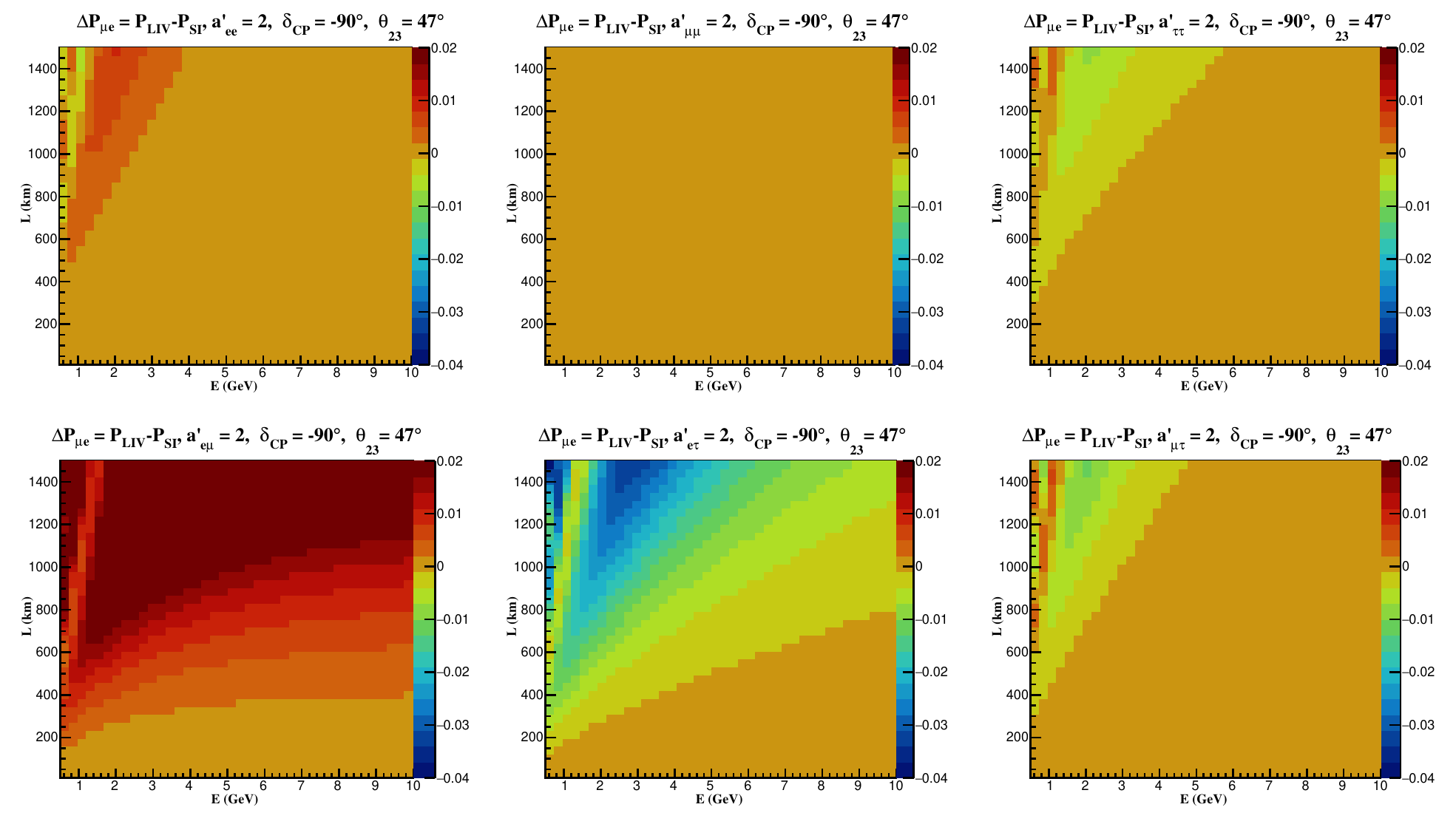}
    \caption{Variation of $\Delta P_{\mu e}=P^{\mu e}_{LIV}-P^{\mu e}_{SI}$ for energy--baseline space with $(a_{ee}, a_{\mu\mu}, a_{\tau\tau})$ in top-panels and $(a_{e\mu}, a_{e\tau}, a_{\mu\tau})$ 
    in bottom panels respectively. Here, we consider the case $a'_{\alpha\beta}= 2$, $\phi_{\alpha\beta}=0^{\circ}$ for all the sub-figures.}
    \label{fig:delPme_E_vs_L}
\end{figure}

\begin{figure}[h]
    \centering
    \includegraphics[width=1\linewidth]{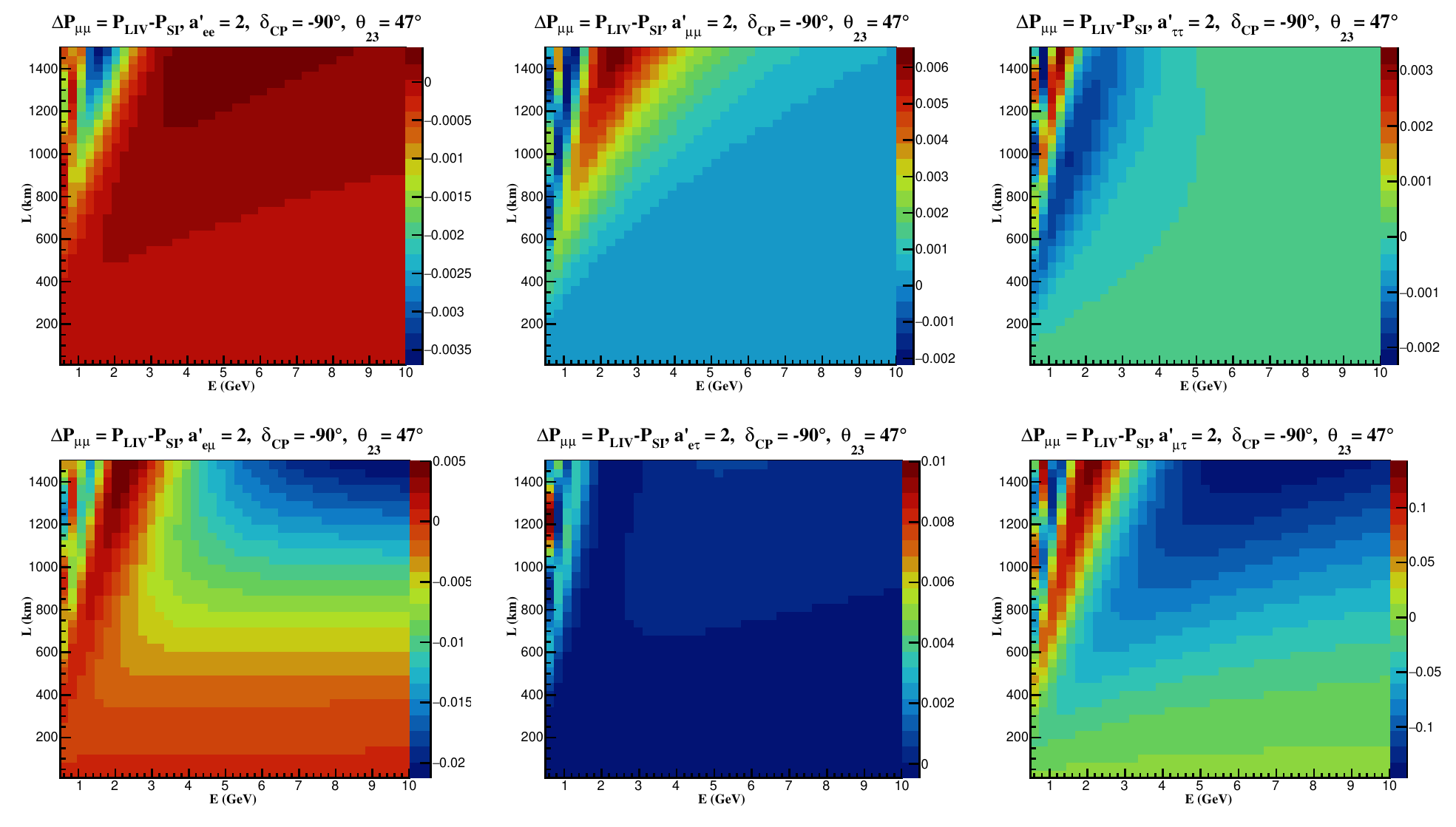}
    \caption{Variation of $\Delta P_{\mu \mu}=P^{\mu e}_{LIV}-P^{\mu e}_{SI}$ on energy--baseline space for $(a_{ee}, a_{\mu\mu}, a_{\tau\tau})$ in top-panels and $(a_{e\mu}, a_{e\tau}, a_{\mu\tau})$ 
    in bottom panels respectively. We consider the case $a'_{\alpha\beta}= 2$, $\phi_{\alpha\beta}=0^{\circ}$.}
    \label{fig:delPmm_E_vs_L}
\end{figure}

We now further explore the effects of LIV on $P_{\mu e}$ and $P_{\mu \mu}$ in the energy-baseline space. For quantifying the effect of LIV, we define a parameter $\Delta P_{\mu e}$ as,
\begin{equation}\label{eq:delP}
\Delta P_{\mu e}=P^{\mu e}_{LIV}-P^{\mu e}_{SI},
\end{equation}
where, $P^{\mu e}_{LIV}$ is the appearance probability in the presence of LIV and $P^{\mu e}_{SI}$ is the appearance probability for the standard case in the absence of LIV. In figure \ref{fig:delPme_E_vs_L}, we show the impact of LIV parameters on the oscillation probabilities as a function of neutrino energy and baseline. The impact of  $a_{ee}$, $a_{\mu\mu}$, $a_{\tau\tau}$  and $a_{e\mu}$, $a_{e\tau}$, $a_{\mu\tau}$ on $\Delta P_{\mu e}$ are shown in top and bottom panels respectively. We take $a'_{\alpha\beta} = 2$ and phase as $\phi_{\alpha\beta}=0^{\circ}$. 
\begin{itemize}
    \item In figure \ref{fig:delPme_E_vs_L}, we observe a nominal effect on $P_{\mu e}$ for $a_{\mu\mu}$. On the other hand, the effect of $a_{ee}$ and $a_{\tau\tau}$ is prominent for longer baselines. 
    \item The variation in $\Delta P_{\mu e}$ is maximum in the presence of $a_{e\mu}$ and $a_{e\tau}$. The effect of $a_{e\mu}$ is notable for L $>$ 180 km.
    \item The effect of $a_{e\tau}$ can be seen for baseline L$>$100 km at lower energies. Also, for $a_{\mu\tau}$ the effect can be seen only at lower energies for the considered value of baselines.
    \item All the observations are in good agreement with the analytical expressions of probabilities and the results from figure \ref{fig:all_Pmue_band}.
\end{itemize}
Similarly, we explore the impact of $a_{\alpha\beta}$ on oscillation channel $P_{\mu\mu}$ by taking $\rm \Delta P_{\mu\mu}= P^{\mu \mu}_{LIV}-P^{\mu \mu}_{SI}$ with varying energy and baseline in figure \ref{fig:delPmm_E_vs_L}. Here, we make the following observations.
 \begin{itemize}
    \item In figure \ref{fig:delPmm_E_vs_L}, we observe that all the diagonal elements have nominal effects on $P_{\mu \mu}$ at lower energies and longer baselines. Here, the effect for $a_{\mu\mu}$ is highest as compared to the other diagonal elements i.e. $a_{ee}$ and $a_{\tau\tau}$.
    \item The effect of off-diagonal element $a_{\mu\tau}$ is highest while the effect from $a_{e\tau}$ is minimal. We see a prominent effect for $a_{e\mu}$ at higher energies and longer baselines.
    \item All the observations are in good agreement with the analytical expressions of probabilities and the results from figure \ref{fig:all_Pmumu_band}.
\end{itemize}

\begin{figure}[h]
    \centering
    \includegraphics[width=1\linewidth]{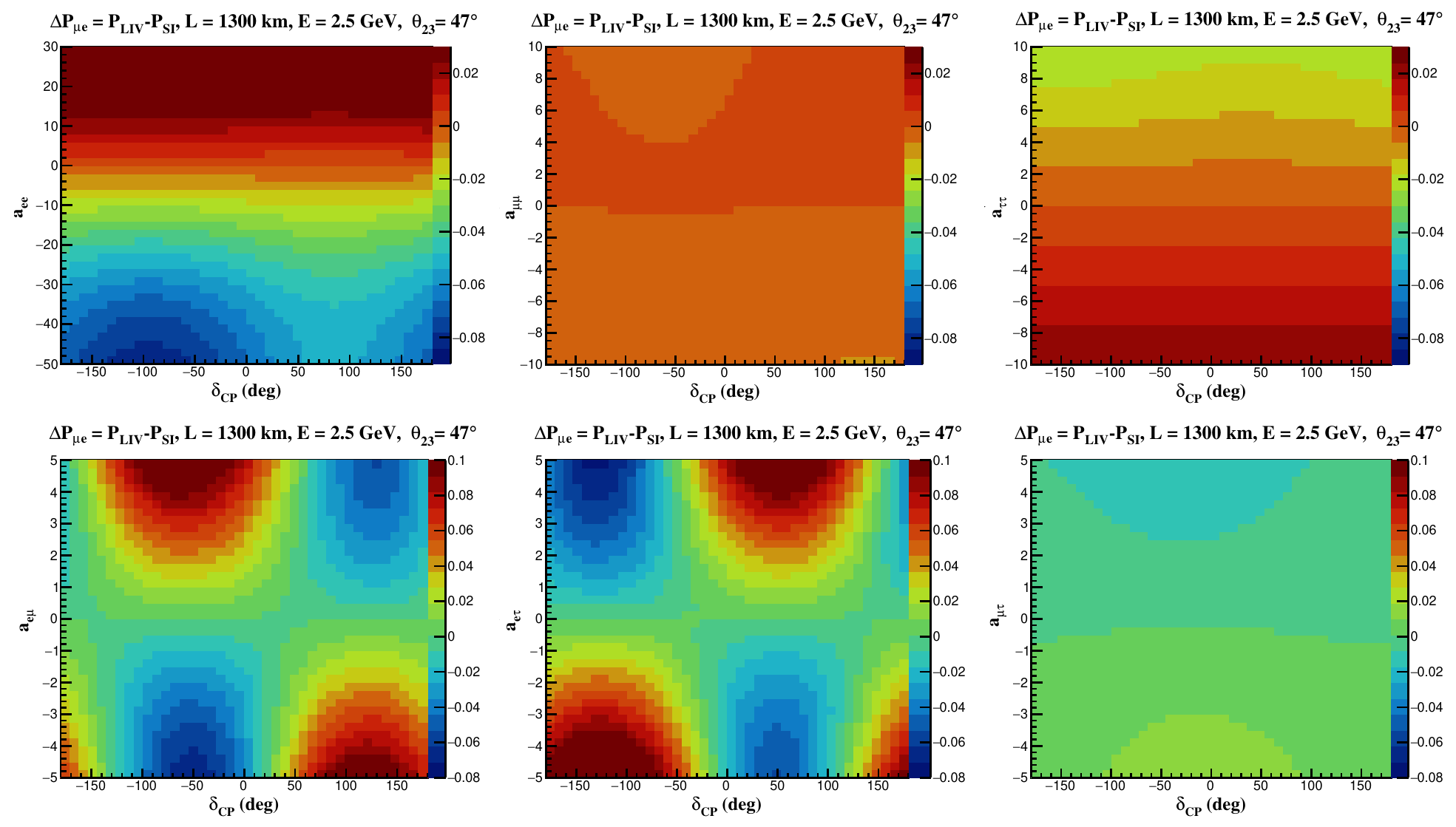}
    \caption{Variation of $\Delta P_{\mu e}=P^{\mu e}_{LIV}-P^{\mu e}_{SI}$ for $a_{\alpha\beta}$ - $\delta_{CP}$ space considering $(a_{ee}, a_{\mu\mu}, a_{\tau\tau})$ in top-panels and $(a_{e\mu}, a_{e\tau}, a_{\mu\tau})$ 
    in bottom panels respectively. We fixed the baseline L = 1300 km, energy at 2.5 GeV.}
    \label{fig:delPme_a_vs_dcp}
\end{figure}

\begin{figure}[h]
    \centering
    \includegraphics[width=1\linewidth]{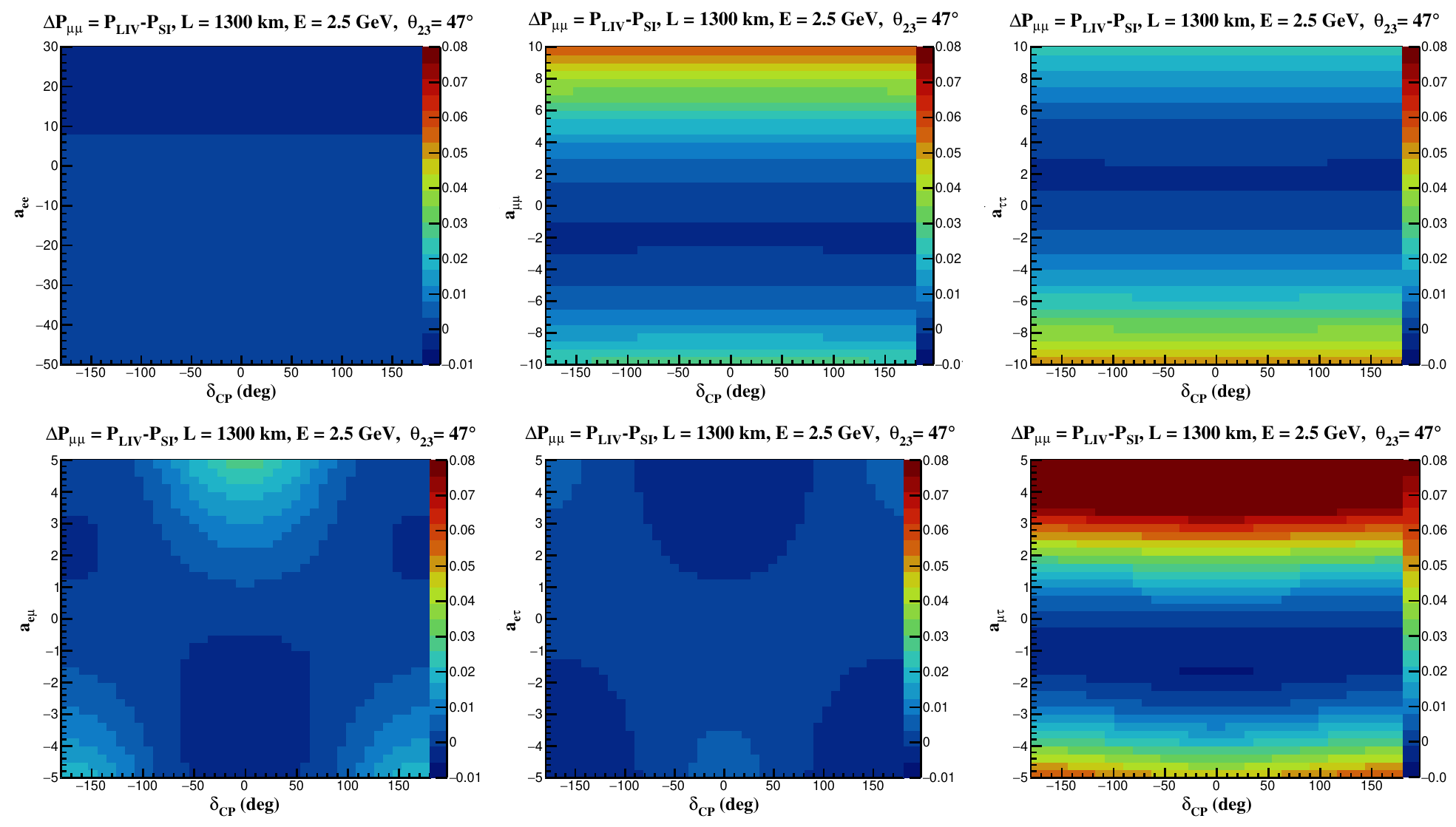}
    \caption{Variation of $\Delta P_{\mu \mu}=P^{\mu e}_{LIV}-P^{\mu e}_{SI}$ for $a_{\alpha\beta}$ - $\delta_{CP}$ space considering $(a_{ee}, a_{\mu\mu}, a_{\tau\tau})$ in top-panels and $(a_{e\mu}, a_{e\tau}, a_{\mu\tau})$ 
    in bottom panels respectively. We fixed the baseline at L=$1300$ km and energy at E = $2.5$ GeV.}
    \label{fig:delPmm_a_vs_dcp}
\end{figure}

From the above study, we conclude that a significant impact of $a_{\alpha\beta}$ can be seen at longer baselines. We note that the region around the DUNE baseline of 1300 km and peak energy $\sim$ 2.5 GeV is promising for probing the effects of LIV. We next explore the $a_{\alpha\beta}$--$\delta_{CP}$ parameter space for both $P_{\mu e}$ and $P_{\mu\mu}$ in the figures \ref{fig:delPme_a_vs_dcp} and \ref{fig:delPmm_a_vs_dcp} respectively. We use the quantity defined in equation \ref{eq:delP} for fixed baseline $L=1300$ km and energy $E=2.5$ GeV.

\begin{itemize}
    \item In figure \ref{fig:delPme_a_vs_dcp}, we observe a suppression (enhancement) of probability for negative (positive) values of $a_{ee}$. For $a_{\tau\tau}$, the probability gets enhanced (suppressed) for negative (positive) values in the complete $\delta_{CP}$ parameter space whereas the effects due to $a_{\mu\mu}$ are very nominal.
    \item In figure \ref{fig:delPme_a_vs_dcp}, we see an enhancement in the negative $\delta_{CP}$ plane and a suppression in the positive plane for positive $a_{e\mu}$. The effects due to $a_{e\tau}$ are opposite in nature to $a_{e\mu}$, where the probability enhances (suppresses) in the positive (negative) $\delta_{CP}$ plane. We observed minimal changes in $\Delta P_{\mu e}$ in the presence of $a_{\mu\tau}$.
\end{itemize}

\begin{itemize}
    \item In figure \ref{fig:delPmm_a_vs_dcp}, the impact of $a_{ee}$ appears to be minimal. With the increase in the strength of $a_{\mu\mu}$ (positive), the probability values get enhanced for the complete $\delta_{CP}$ parameter space. Whereas, for $a_{\tau\tau}$, the enhancement can be seen for negative values in $\delta_{CP}\in[-\pi,\pi]$. 
    \item In figure \ref{fig:delPmm_a_vs_dcp}, for $a_{e\mu}$ and $a_{e\tau}$, we see nominal effects on the probability. In the presence of $a_{\mu\tau}$, we see a prominent effect with the increase in the strength for the complete $\delta_{CP}$ space.
\end{itemize}

\begin{figure}[h]
    \centering
    \includegraphics[width=1\linewidth]{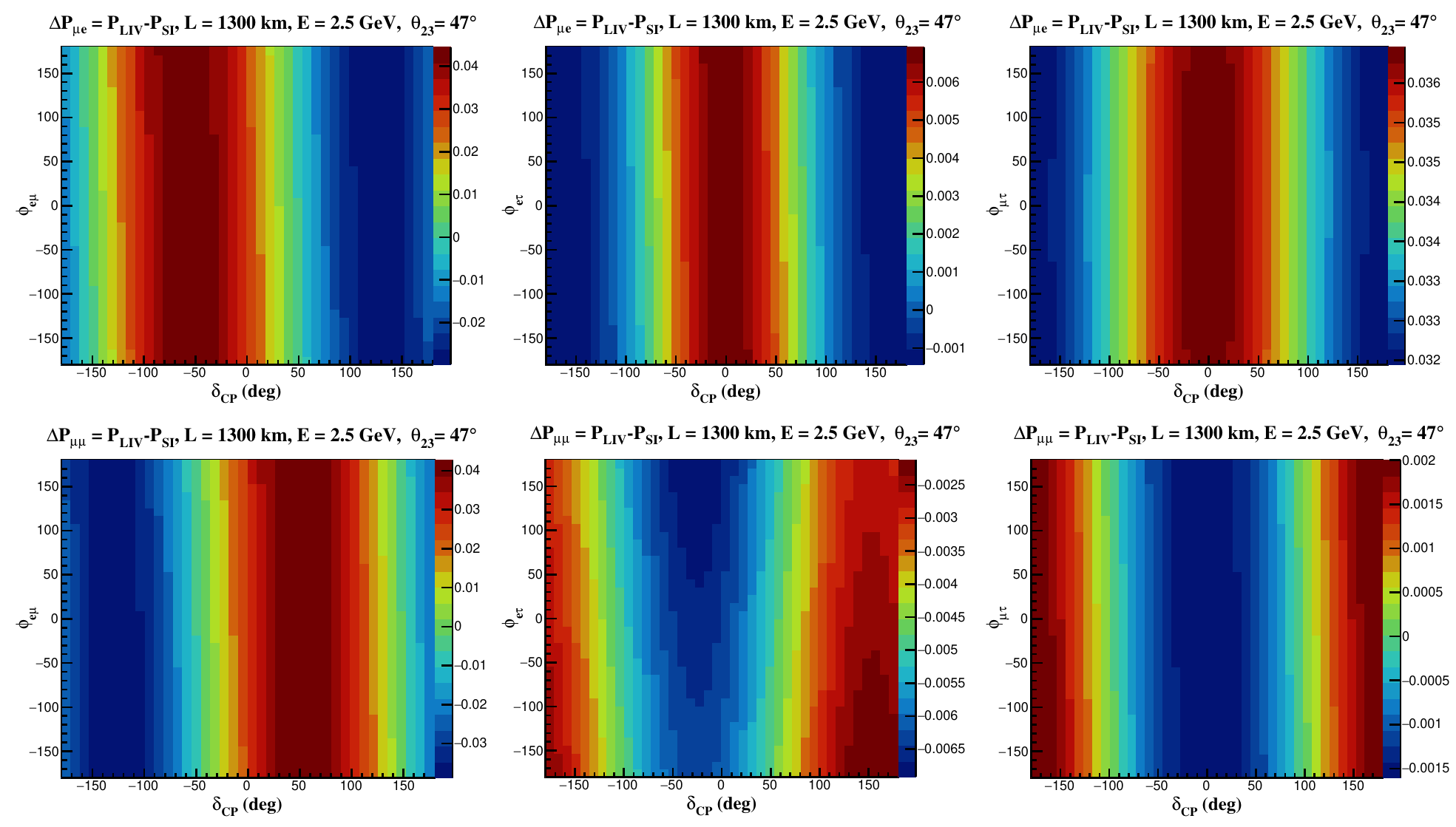}
    \caption{Variation of $\Delta P_{\mu e}$ (top-panels) and $\Delta P_{\mu \mu}$ (bottom-panels) for $\phi_{\alpha\beta}$ - $\delta_{CP}$ space considering $(a_{e\mu}, a_{e\tau}, a_{\mu\tau})$ 
    in left, middle and right panels respectively. We fixed the $a_{\alpha\beta}$=2, baseline $L = 1300$km, $E = 2.5$ GeV.}
    \label{fig:delPme_phi_vs_dcp}
\end{figure}

We also explore $\phi_{\alpha\beta}$ - $\delta_{CP}$ parameter space for both $P_{\mu e}$ and $P_{\mu\mu}$ in the figure \ref{fig:delPme_phi_vs_dcp} where the top (bottom) panels show the effects for $P_{\mu e}$ ($P_{\mu\mu}$) taking the off-diagonal elements $(a_{e\mu},a_{e\tau},a_{\mu\tau})$ respectively. Here, we use the quantity defined in equation \ref{eq:delP} for fixed baseline L=$1300$ km and energy E=$2.5$ GeV.

\begin{itemize}
    \item In figure \ref{fig:delPme_phi_vs_dcp}, we observe an enhancement in $\Delta P_{\mu e}$ for the negative $\delta_{CP}$ plane whereas a suppression in the positive $\delta_{CP}$ plane for the complete $\phi$ parameter space due to the presence of $a_{e\mu}$.
    \item For $a_{e\tau}$ and $a_{\mu\tau}$, we see maximum $\Delta P_{\mu e}$ enhancement around $\delta_{CP}=0$ and a suppression for non-zero value of $\delta_{CP}$ in both negative and positive plane.
    \item For $a_{e\mu}$, we see an enhancement in $\Delta P_{\mu \mu}$ in the $\delta_{CP}$ region $[-20^{\circ},130^{\circ}]$.
    \item In the case of $a_{e\tau}$, a suppression of $\Delta P_{\mu \mu}$ is observed in the complete $\delta_{CP}$ region which peaks around $\delta_{CP}=-30^{\circ}$. In the presence of $a_{\mu\tau}$, the suppression can be observed in the $\delta_{CP}$ region $[-80^{\circ},80^{\circ}]$.
\end{itemize}

For long-baseline neutrino experiments, the oscillation channels $\nu_{\mu}\rightarrow\nu_{e}$ and $\nu_{\mu}\rightarrow\nu_{\mu}$ are used to study the octant, Dirac CP phase, and mass hierarchy sensitivities. The $\nu$-oscillation channel $\nu_{\mu}\rightarrow\nu_{e}$ is most sensitive in all the long-baseline experiments and helps in probing CP-violation.
As shown in this section, we have explored the effects of various LIV parameters at the probability level. The sensitive parameters affecting $P_{\mu e}$ are $a_{e\mu}$ and $a_{e\tau}$, while for $P_{\mu\mu}$ is $a_{\mu\tau}$. We have also observed that the presence of LIV phases can significantly modify the oscillation probabilities. In figure \ref{fig:Pmue_vs_dcp} and \ref{fig:Pmumu_vs_dcp}, we explored the effects of $a_{\alpha\beta}$ on the probability channels with varying $\delta_{CP}$. The presence of the off-diagonal phase modifies the probabilities for the whole $\delta_{CP}$ parameter space significantly for $a_{e\mu}$, $a_{e\tau}$ and $a_{\mu\tau}$. Figure \ref{fig:delPme_E_vs_L} and \ref{fig:delPmm_E_vs_L} indicates that a significant impact of $a_{\alpha\beta}$ can be seen at longer baselines. A scan of $a_{\alpha\beta}-\delta_{CP}$ parameter space in figure \ref{fig:delPme_a_vs_dcp} and \ref{fig:delPmm_a_vs_dcp} also implies a major contribution to the changes in probabilities from $a_{e\mu}$, $a_{e\tau}$ and $a_{\mu\tau}$. In figure \ref{fig:delPme_phi_vs_dcp}, we have explored the $\phi_{\alpha\beta}-\delta_{CP}$ space which shows possible effects of off-diagonal phases on the probabilities for varying $\delta_{CP}$.

Motivated by the effect of the LIV parameters at the probability level, we explore next, the impact at the $\chi^2$ level. We particularly consider the most sensitive parameters i.e. $(a_{e\mu},a_{e\tau},a_{\mu\tau})$ and study the $\chi^2$ sensitivities.  We have simulated the upcoming long baseline neutrino experiment DUNE as the test case. In this work, we have examined the implications of CPT violating Lorentz violation effects on the appearance and disappearance probability channels by considering only one LIV parameter at a time. In the following subsection \ref{sec:Simulation}, we describe the experimental specifications of DUNE and the inputs for the simulation.
\subsection{DUNE: Experimental details and inputs for simulation}\label{sec:Simulation}
The Deep Underground Neutrino Experiment (DUNE) \cite{DUNE:2015lol,DUNE:2016hlj,DUNE:2020txw} is a future long-baseline neutrino beam experiment that will explore the neutrino mixing parameters using a high power muon neutrino (anti-neutrino) beam. It consists of a Near detector (ND) situated at Fermilab and a Far detector (FD) to be located at a distance of 1300 km away at Sanford Underground Research Facility (SURF), South Dakota, USA. This makes DUNE a very long baseline of 1300 km. The FD is a fiducial 40 kton liquid argon time projection chamber (LArTPC) located underground to eliminate background sources. The powerful muon neutrino (anti-neutrino) beam to be used is intended to have a power of 1.2 MW and protons-on-target of approximately 1.1$\times10^{21}$ POT/yr. The beam is expected to peak at an energy of 2.5 GeV. The primary objectives of the experiment are to precisely establish the neutrino mixing parameters, investigate matter-antimatter asymmetry via  Charge-Parity (CP) symmetry violation and identify the true neutrino mass hierarchy. This experiment will provide unmatched precision in establishing neutrino mixing parameters. It will be able to set stringent constraints on Lorentz and CPT violation with the neutrino sector and test theoretical foundations of quantum field theory. 
\begin{table}[h]
    \centering
    \begin{tabular}{|c|c|c|c|c|}
    \hline 
    Detector Details (LArTPC, 35kton) & \multicolumn{2}{c|}{Normalization error} & \multicolumn{2}{c|}{Energy Calibration error}\tabularnewline
    \hline 
    Runtime ($5\nu+5\bar{\nu}$) yr & Signal & Background & Signal & Background\tabularnewline
    \hline 
    $\varepsilon_{app}=80\%, \varepsilon_{dis}=85\%,$ & $\nu_{e}:5\%$ & $\nu_{e}:10\%$ & $\nu_{e}:5\%$ & $\nu_{e}:5\%$\tabularnewline
    \hline 
    $R_{e}=0.115/\sqrt{E}$,$R_{\mu}=0.2/\sqrt{E}$ & $\nu_{\mu}:5\%$ & $\nu_{\mu}:10\%$ & $\nu_{\mu}:5\%$ & $\nu_{\mu}:5\%$\tabularnewline
    \hline 
    \end{tabular}
    \caption{Specifications of detector efficiency, resolution, and systematic uncertainties for DUNE}
    \label{tab:DUNE_spec_table}
\end{table}
In table \ref{tab:DUNE_spec_table}, $\varepsilon_{app}$ and $\varepsilon_{dis}$ are the signal efficiencies for $\nu_{e}^{CC}$ and $\nu_{\mu}^{CC}$  whereas, $R_{e}$ and $R_{\mu}$ represent the energy resolution for $\nu_{e}^{CC}$ and $\nu_{\mu}^{CC}$ respectively.

For simulation purpose, we use the General Long Baseline Experiment Simulator (GLoBES) \cite{Huber:2004ka,HUBER2007439} which is a simulation package extensively used for simulation of long-baseline neutrino experiments. With a run-time of 5 years in neutrino mode and 5 years in anti-neutrino mode for DUNE, we take into account a total exposure of 35$\times10^{22}$ kt-POT-yr in our simulation. For both modes of operation, the background normalization error is taken to be 10$\%$ and the signal normalization for neutrino (anti-neutrino) mode is taken to be 2$\%$ (5$\%$).

To distinguish between the standard interaction and the effects due to the presence of LIV, we define the statistical $\chi^2$ as, 
\begin{equation}
\label{eq:chisq}
\chi^2 \equiv  \min_{\eta}  \sum_{i} \sum_{j}
\frac{\left[N_{true}^{i,j} - N_{test}^{i,j} \right]^2 }{N_{true}^{i,j}},
\end{equation}
where, $N_{true}^{i,j}$ and $N_{test}^{i,j}$ are the number of true and test events in the $\{i,j\}$-th bin respectively. 
All the systematic errors are incorporated using the pull method described in the \cite{Huber:2004ka,Fogli:2002pt}. They can be introduced by additional variables $\zeta_{k}$ called nuisance parameters.
\begin{equation}
    \chi_{pull}^{2}=\underset{\zeta_{j}}{min}\left(\chi^{2}+\sum_{i=1}^{k}\frac{\zeta_{i}^{2}}{\sigma_{\zeta_{i}}^{2}}\right),
\end{equation}
where, $\sigma_{\zeta_{k}}^{2}$ is the systematical error/uncertainty of $\zeta_{k}^{th}$ nuisance parameter. The total measure of statistical significance can be obtained by minimizing over all the systematic errors considered.


\section{Results and Discussions}\label{sec:results}

As discussed in section \ref{sec:delPmue}, we see that a significant contribution to the oscillation probabilities arises due to the presence of LIV parameters. 
The impact of $a_{e\mu}$ and $a_{e\tau}$ are significant on $P_{\mu e}$ whereas, the impact of $a_{\mu\tau}$ is notable on $P_{\mu\mu}$. We also observe a prominent impact of LIV parameters on $\delta_{CP}$ space. It may also be noted that the presence of the off-diagonal phases may also alter the probabilities. Hence, the presence of LIV can affect the CP-measurement sensitivity. In this work, we have primarily focused on exploring the impact of CPT-violating LIV parameters (in particular, $a_{e\mu}$, $a_{e\tau}$, and $a_{\mu\tau}$) on the CP-measurement sensitivity at DUNE. We also examine the CP-precision sensitivity in the presence of these parameters. In \ref{sec:sens_LIV}, we explore the sensitivity of DUNE towards constraining the LIV parameters. We then discuss the CP-violation and CP-precision sensitivity in the subsections \ref{sec:CPV_sens} and \ref{sec:CPPrec_sens} respectively.

\subsection{Constraining the LIV parameters with DUNE} \label{sec:sens_LIV}
\begin{figure}[h]
\centering
\begin{minipage}{.49\textwidth}
  \centering
  \includegraphics[width=1.1\linewidth,  height = 6cm]{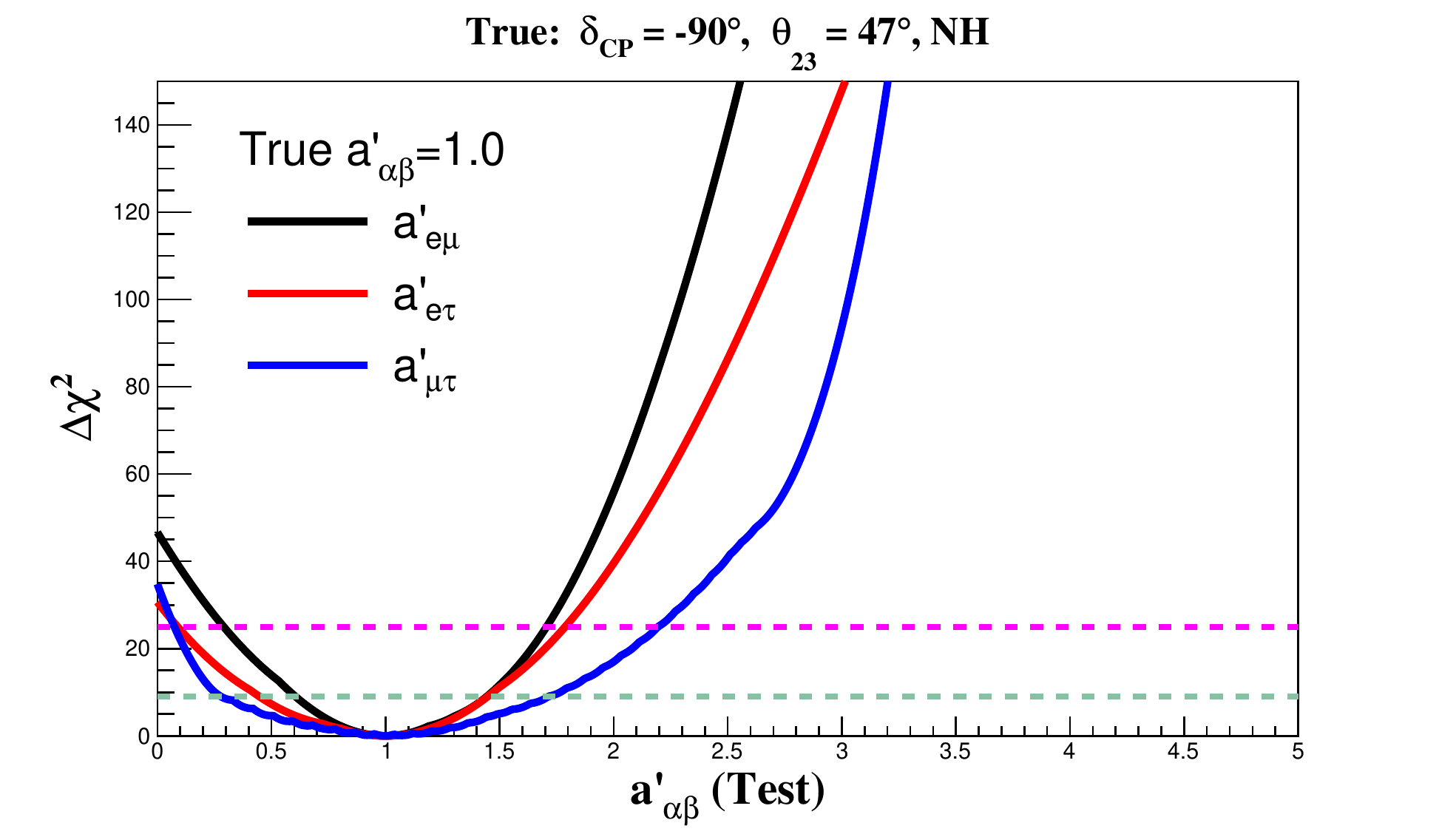}
\end{minipage}
\caption{Sensitivity of DUNE in constraining $a_{\alpha\beta}$. We show the results for the elements $(a_{e\mu}$, $a_{e\tau}$, $a_{\mu\tau})$ where we take $a'_{\alpha\beta}=1.0$, $\phi_{\alpha\beta}=0^{\circ}$, $\delta_{CP}=-\pi/2$ and $\theta_{23}=47^{\circ}$. The parameter $a'_{\alpha\beta}$ represents the values of $a_{\alpha\beta}$ as expressed in equation \ref{eq:prime_offdiagonal}. The dashed magenta and green lines represent $5\sigma$ and $3\sigma$ CL respectively. The black, red and blue solid lines represent the case with the parameters $a_{e\mu}$, $a_{e\tau}$ and $a_{\mu\tau}$ respectively.}
\label{fig:liv_testing}
\end{figure}

We present the sensitivity of DUNE towards constraining $a_{\alpha\beta}$ in figure \ref{fig:liv_testing}. The results for LIV parameters $(a_{e\mu}$, $a_{e\tau}$, $a_{\mu\tau})$ are included in the figure \ref{fig:liv_testing}. The true values of $a'_{\alpha\beta}$ are fixed at $1.0$ and the test values are varied in the range $[0,5]$. The black, red and blue solid lines represent the case with the parameters $(a_{e\mu}$, $a_{e\tau}$, $a_{\mu\tau})$. The magenta and green dashed line represent the $5\sigma$ and $3\sigma$ confidence levels. 
\begin{figure}[h]
    \centering
    \includegraphics[width=0.6\textwidth,  height = 6cm]{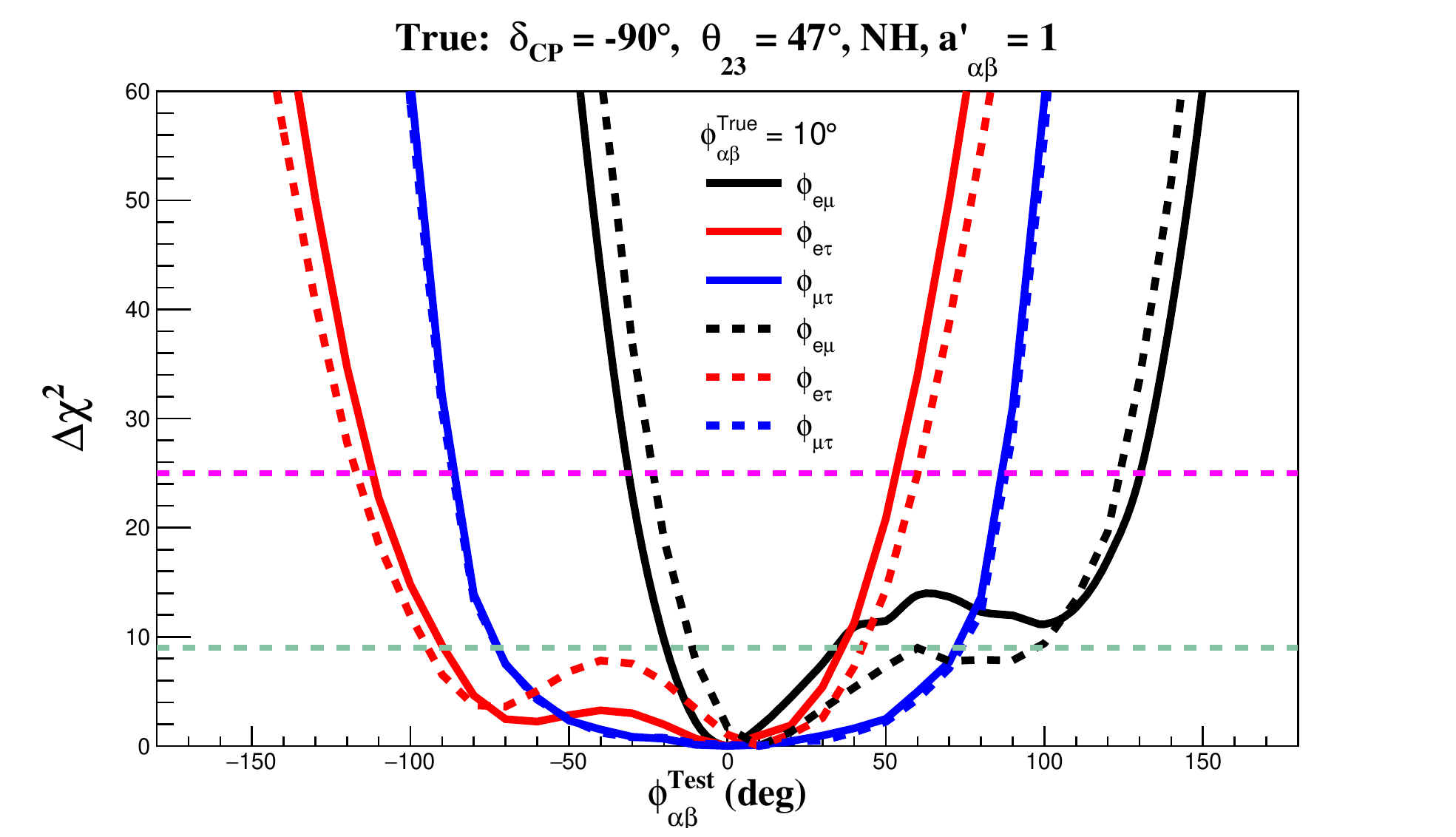}
    \caption{Sensitivity of DUNE for the LIV phases $\phi_{\alpha\beta}$ with $a'_{\alpha\beta}=1.0$, $\delta_{CP}=-\pi/2$, $\theta_{23}=47^{\circ}$ and $\phi_{\alpha\beta}=10^{\circ}$. The parameter $a'_{\alpha\beta}$ represents the values of $a_{\alpha\beta}$ as expressed in equation  \ref{eq:prime_offdiagonal}. The solid lines represent the case with $\phi_{\alpha\beta}=0^{\circ}$, whereas the dashed lines represents the case with $\phi_{\alpha\beta}=10^{\circ}$. The dashed magenta and green line represent $5\sigma$ and $3\sigma$ CL respectively.}
    \label{fig:liv_w_phase_testing}
\end{figure}
In figure \ref{fig:liv_testing}, the $\Delta \chi^{2}$ sensitivity of off-diagonal parameters $(a_{e\mu},a_{e\tau},a_{\mu\tau})$ are shown. The LIV parameter $a_{e\mu}$ is best constrained out of all the off-diagonal parameters.
In figure \ref{fig:liv_w_phase_testing}, we present the $\Delta \chi^{2}$ sensitivities of DUNE for non-zero LIV phases $\phi_{e\mu}$, $ \phi_{e\tau}$ and $\phi_{\mu\tau}$.
The presence of the phases in the LIV elements can bring in more degeneracy in terms of the measurement of $\delta_{CP}$. It may also affect the sensitivity of DUNE for $\delta_{CP}$-measurement. Here we have marginalized over $\delta_{CP}$ in the range $[-\pi,\pi]$. The solid (dashed) lines represent the case where $\phi_{\alpha\beta}=0^{\circ}$ ($\phi_{\alpha\beta}=10^{\circ}$). The dashed magenta and green line represent $5\sigma$ and $3\sigma$ CL respectively. For the chosen set of mixing parameters, we observe a reasonable constraining capability. 

\subsection{The CPV Sensitivity of DUNE in presence of LIV}\label{sec:CPV_sens}
We examine the impact of LIV parameters on the sensitivity of DUNE to determine the CP-violation. The CPV sensitivity of the experiment is the ability of the experiment to exclude the CP-conserving values i.e $\delta_{CP}=0,\pm\pi$.
We define the $\Delta \chi^2$ sensitivity for CPV as follows,
\begin{equation}
{\Delta \chi}^{2}_{\rm CPV}~(\delta^{\rm true}_{\rm CP}) = {\rm min}~\left[\chi^2~(\delta^\text{true}_{CP},\delta^\text{test}_{CP}=0),~\chi^2 (\delta^\text{true}_{CP},\delta^\text{test}_{CP}=\pm \pi)\right ].
\end{equation}

We quantify the CPV sensitivity by calculating the minimum of the defined $\Delta\chi^{2}_{CPV}$ for all true $\delta_{CP}$ values and this provides a measure of CP-violation for the complete $\delta_{CP}$ parameter space. We have calculated ${\Delta \chi}^{2}_{\rm CPV}$ by varying $\delta_{CP}^{true}$ in the range $[-\pi,\pi]$ after marginalizing over the mixing parameters $\theta_{23}$ and mass squared splitting  $\Delta m_{31}^{2}$. We have assumed the higher octant (HO) as the true octant and the normal hierarchy (NH) as the true hierarchy in this study, unless otherwise mentioned. The values of mixing parameter listed in table \ref{tab:parameters} were used to produce the $\Delta \chi^{2}$ sensitivity  plots. The significance of CPV can be obtained by using $\sigma=\sqrt{{\Delta \chi}^{2}}$, where 5$\sigma$ (3$\sigma$) corresponds to the line at $\sqrt{{\Delta \chi}^{2}}$ = 25 (9) respectively.

\begin{figure}[t]
    \centering
    \begin{minipage}{0.5\textwidth}
      \centering
      \includegraphics[width=1.1\textwidth,  height = 5.5cm]{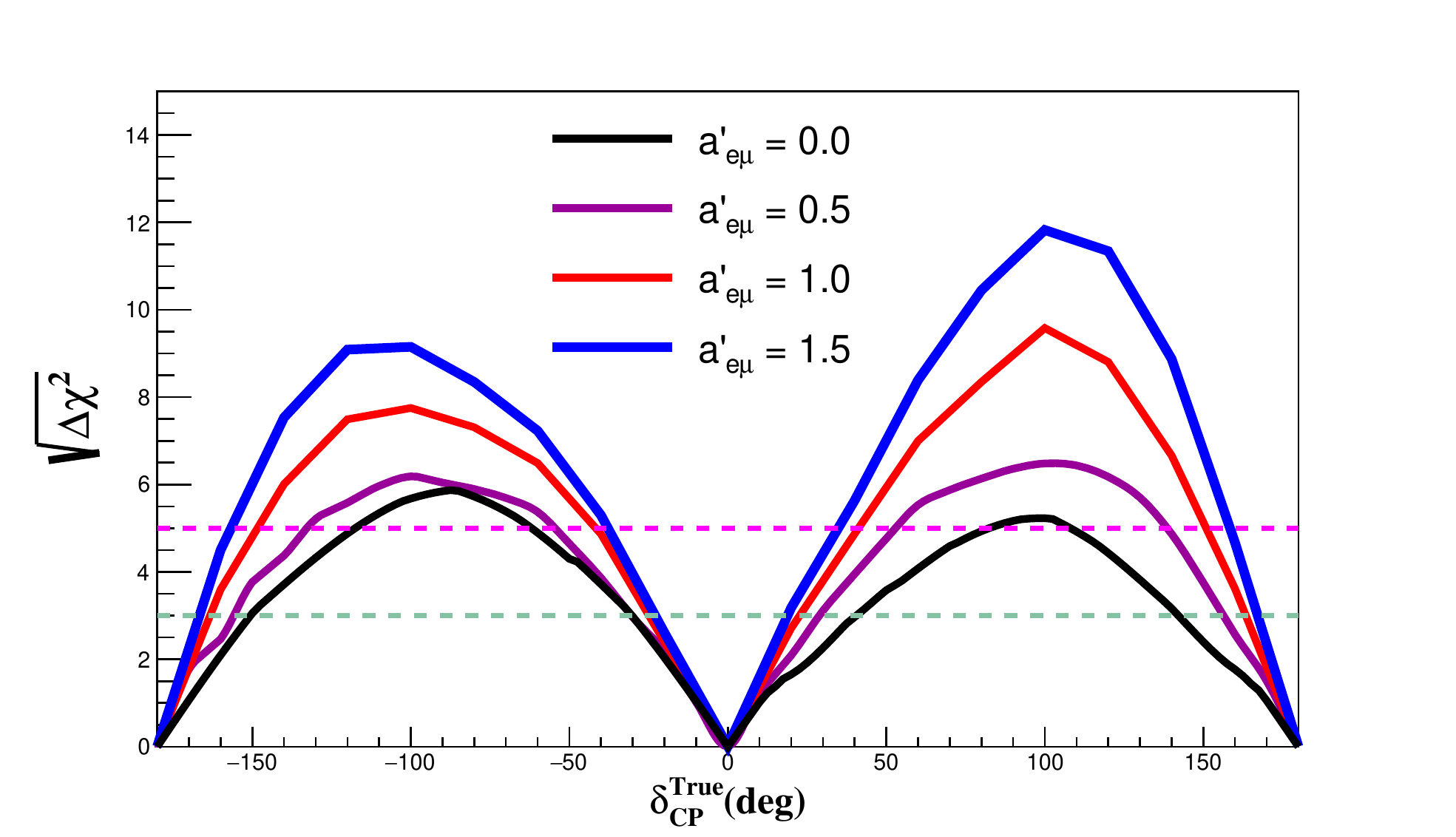}
    \end{minipage}%
    \begin{minipage}{.5\textwidth}
      \centering
      \includegraphics[width=1.1\linewidth,  height = 5.5cm]{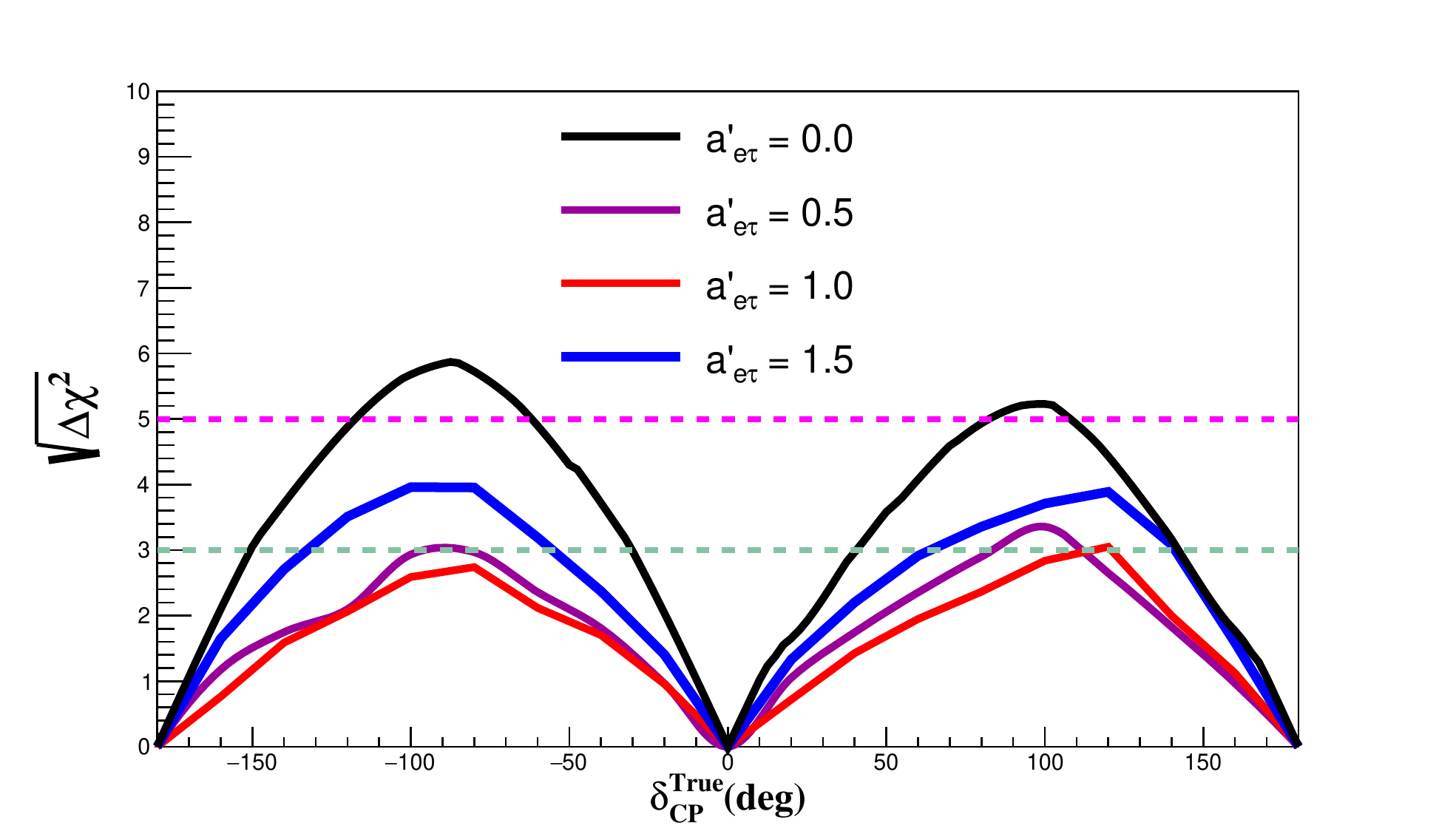}
    \end{minipage}
    \begin{minipage}{.5\textwidth}
      \centering
      \includegraphics[width=1.1\linewidth,  height = 5.5cm]{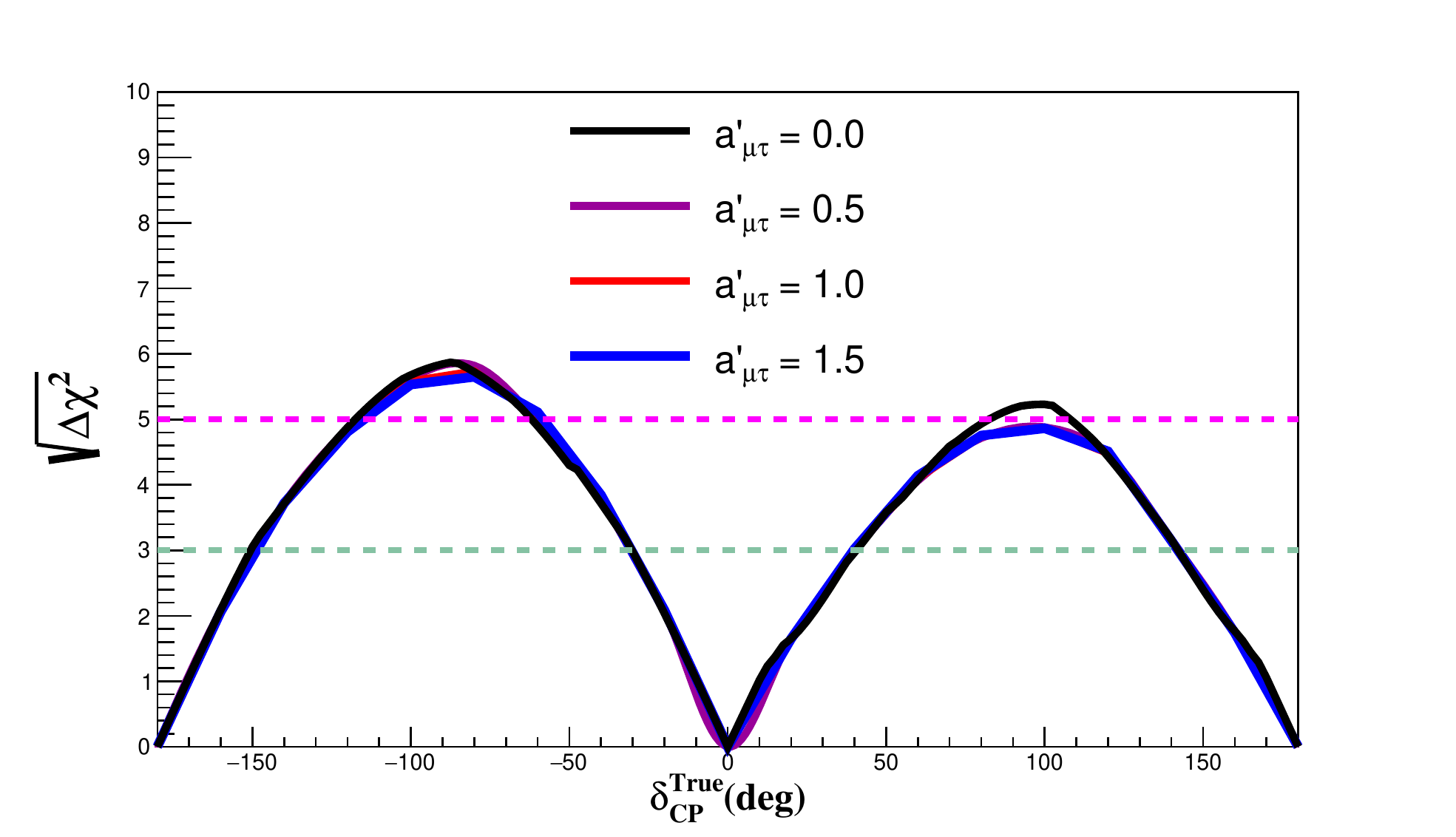}
    \end{minipage}
    \caption{The CPV sensitivity of DUNE for chosen values of $a_{e\mu}$ (top-left panel), $a_{e\tau}$ (top-right panel) and $a_{\mu\tau}$ (bottom panel). We considered true $\delta_{CP}=-\pi/2$, $\theta_{23}=47^{\circ}$ and $\phi_{\alpha\beta}=0$. The parameter $a'_{\alpha\beta}$ represents the values of $a_{\alpha\beta}$ as expressed in equation  \ref{eq:prime_offdiagonal}. The black solid line represents the standard case along with the dashed magenta (green) line representing the 5$\sigma$ (3$\sigma$) CL.}
    \label{fig:chi2_liv}
\end{figure}

We present the CPV sensitivity of DUNE for chosen values of $a_{\alpha\beta}$ in the figure \ref{fig:chi2_liv}, which includes additional marginalization with test value of $a'_{\alpha\beta}$ in the range $[0,2]$. The CPV sensitivities of DUNE for $a_{e\mu}$, $a_{e\tau}$ and $a_{\mu\tau}$ are shown in the top-left, top-right and bottom panels respectively. In figure \ref{fig:chi2_liv}, the black lines represent the standard case with no LIV and other coloured lines represent the cases with non-zero LIV parameters. The magenta and green dashed line represents the $5\sigma$ and $3\sigma$ CL range respectively. 
We note the following,

\begin{itemize}
    \item The increase in the strength of the LIV parameter $a_{e\mu}$ (top-left panel) results in an enhancement of the CPV sensitivity for the complete $\delta_{CP}$ space. The enhancement is higher in the positive region of $\delta_{CP}$ as compared to the negative region of $\delta_{CP}$.
        
    \item In the presence of $a_{e\tau}$ (top-right panel), the sensitivity is suppressed as compared to the standard case. We notice an irregular pattern with an increase in the strength of $a_{e\tau}$, e.g., the sensitivity for $a'_{e\tau}$ = 1.0 appears to be the most suppressed among the chosen values (even marginally more suppressed than that for $a'_{e\tau}$ = 0.5), while that for $a'_{e\tau} = 1.5$  is the least suppressed. This irregular pattern  majorly relates to the irregular pattern that we observe in the appearance probabilities in figure ~\ref{fig:Pmue_vs_dcp}.
    
    \item For the presence of $a_{\mu\mu}$ (bottom), the effect on sensitivity is minimal with slight suppression at the points of maxima.
\end{itemize}

\begin{figure}[t]
\centering
\begin{minipage}{0.49\textwidth}
  \centering
  \includegraphics[width=1.1\textwidth, height = 5.5cm]{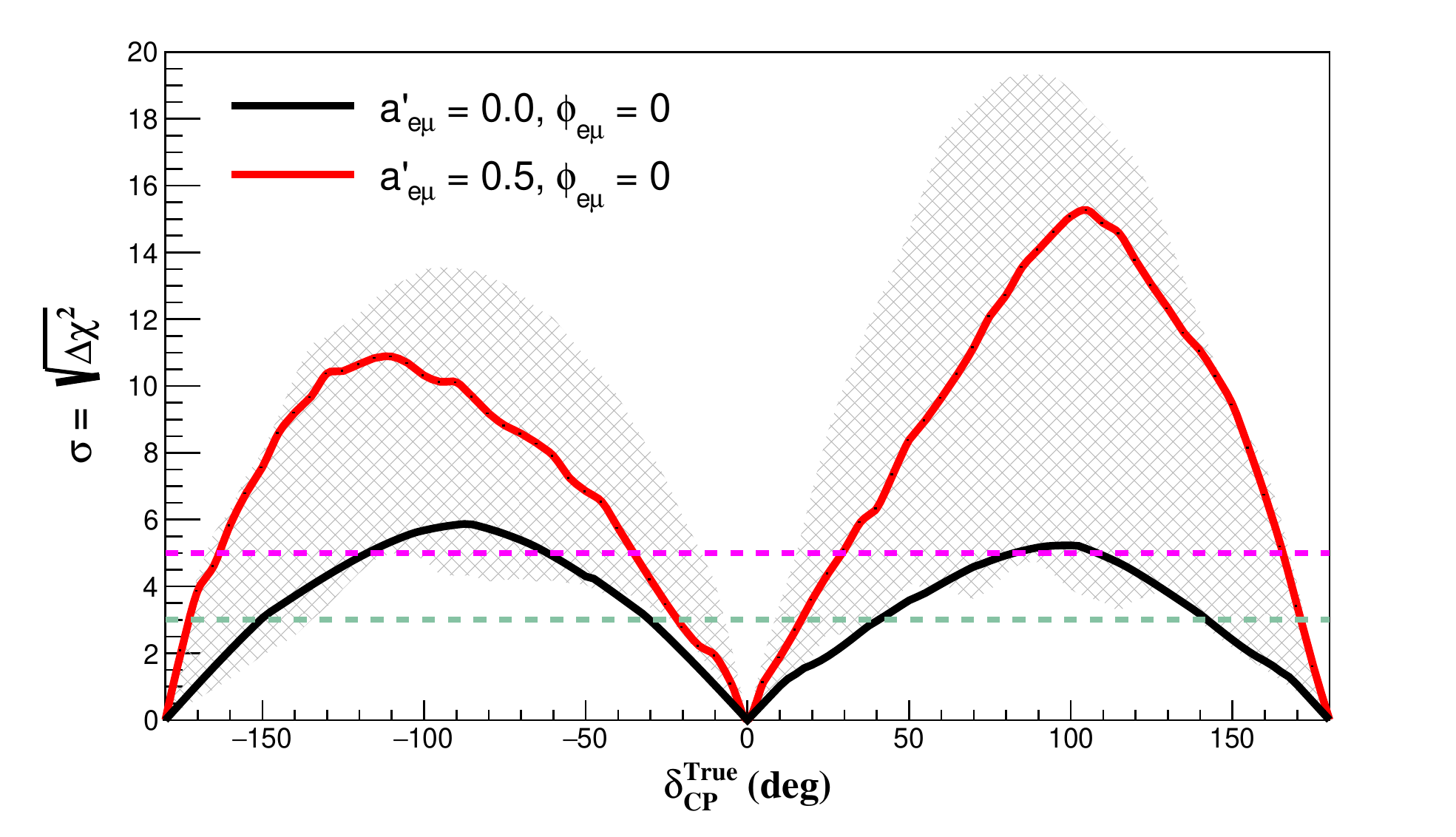}
\end{minipage}
\begin{minipage}{0.49\textwidth}
  \centering
  \includegraphics[width=1.1\textwidth, height = 5.5cm]{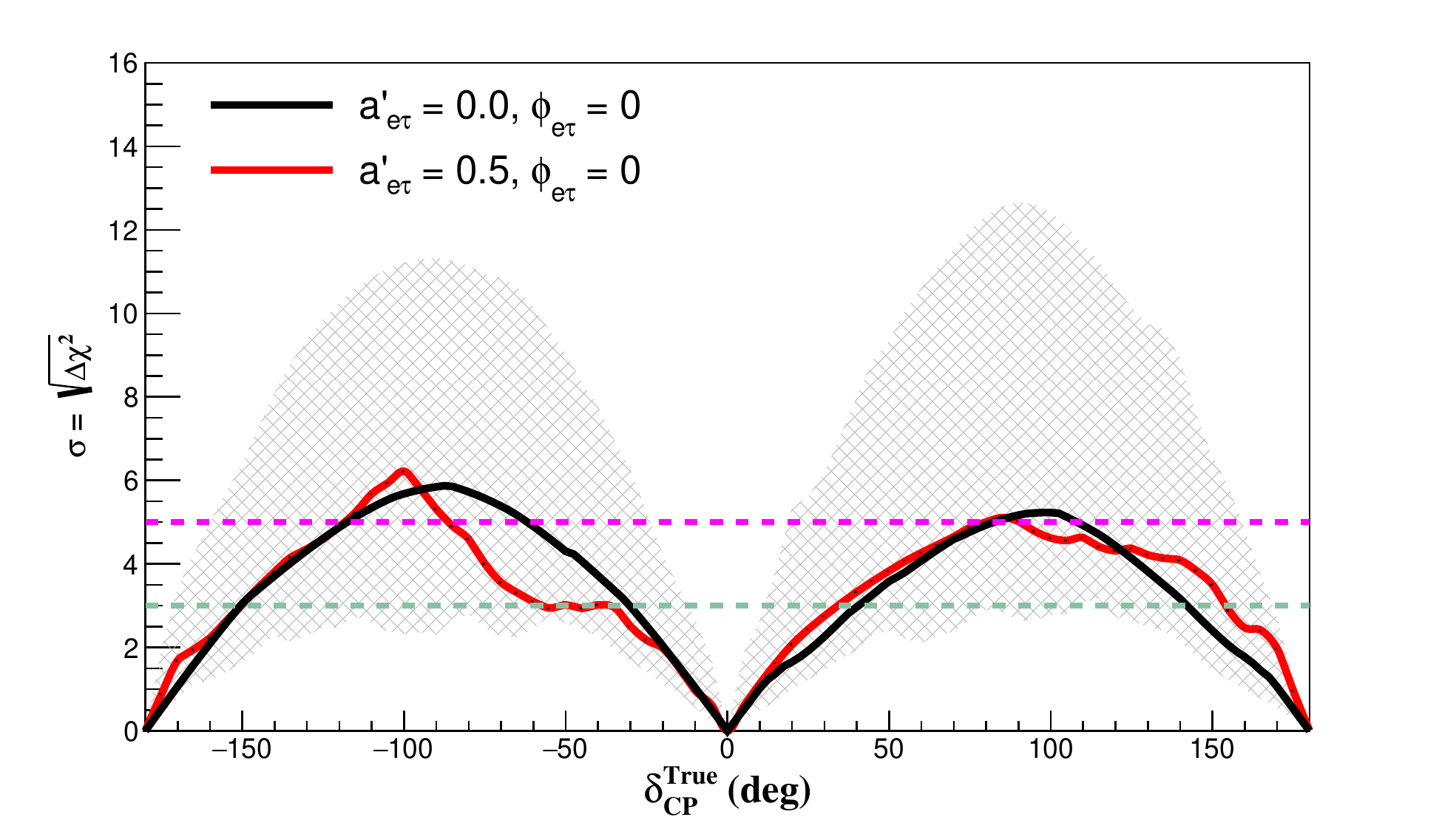}
\end{minipage}
\begin{minipage}{0.49\textwidth}
  \centering
  \includegraphics[width=1.1\textwidth, height = 5.5cm]{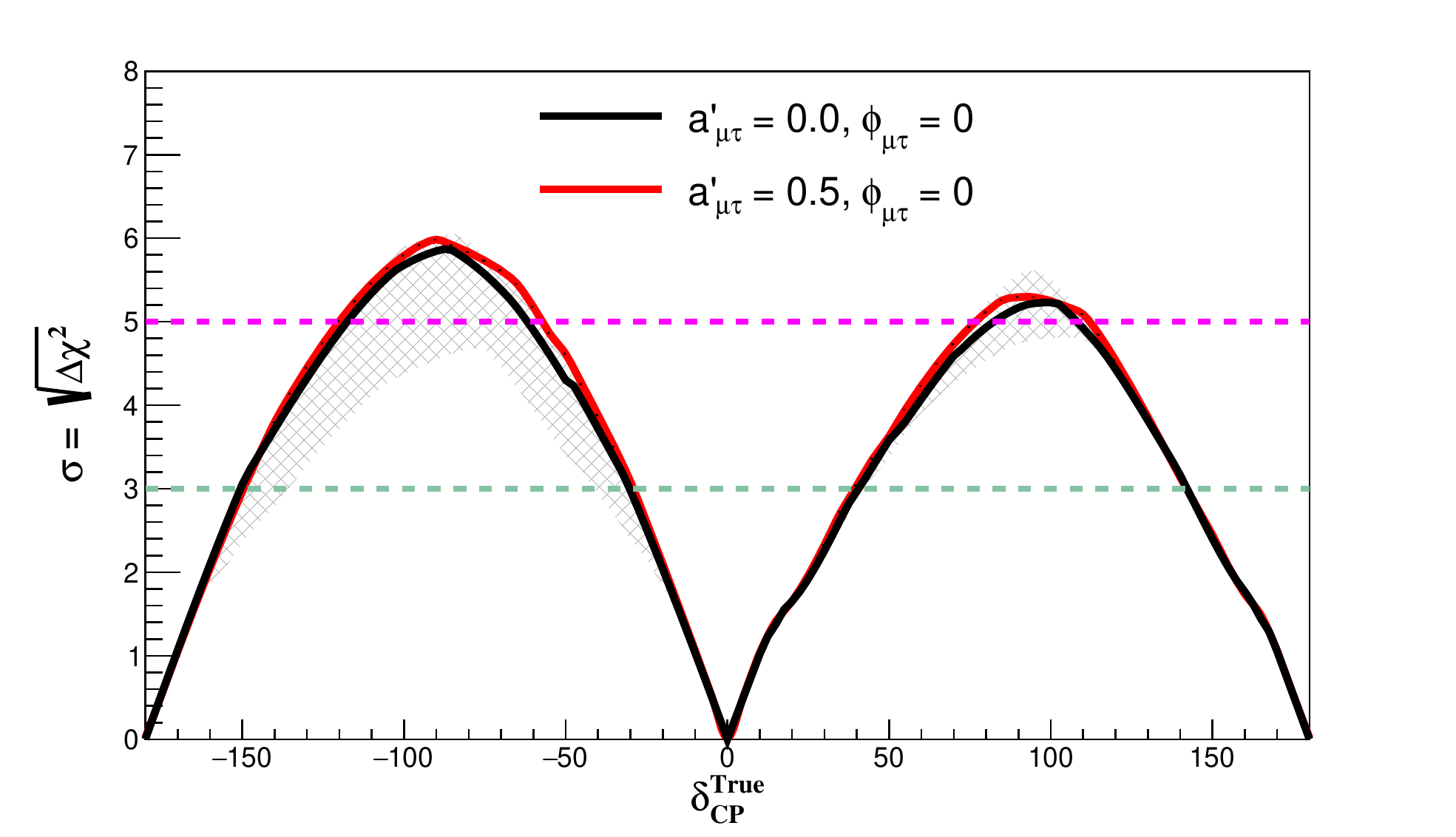}
\end{minipage}
\caption{The CPV sensitivity of DUNE in presence of the off-diagonal LIV parameters $a_{e\mu}$ (top-left panel), $a_{e\tau}$ (top-right panel) and $a_{\mu\tau}$ (bottom panel). In every subfigure, the black solid line represents the standard case with no LIV effects whereas the red solid line represents the case with $|a'_{\alpha\beta}|$ = 0.5 and $\phi_{\alpha\beta}=0$. The shaded region represents the sensitivities corresponding to varied LIV phases $\phi_{\alpha\beta}$ in the range [$-\pi$,$\pi$] and fixed $|a'_{\alpha\beta}|$ = 0.5. The parameter $a'_{\alpha\beta}$ represents the values of $a_{\alpha\beta}$ as expressed in equation \ref{eq:prime_offdiagonal}. The dashed magenta and green line represent $5\sigma$ and $3\sigma$ CL respectively.}
\label{fig:cp_sensitivity}
\end{figure}
\vspace{0.1cm}

In figure \ref{fig:cp_sensitivity}, we show the effects of the LIV phases on the CPV sensitivities. For calculating $\Delta \chi^2$, we have marginalized over the parameters $\theta_{23}$, $\Delta m_{31}^{2}$, $a_{\alpha \beta}$ and $\phi_{\alpha\beta}$ over the 3$\sigma$ ranges as mentioned in Table~\ref{tab:parameters}. The magenta and green dashed line represents the $5\sigma$ and $3\sigma$ CL respectively. The black solid lines represent the standard case and the red solid lines in top-left, top-right and bottom panel represent the LIV sensitivities for non-zero values of $a_{e\mu}$, $a_{e\tau}$ and $a_{\mu\tau}$ respectively. The shaded region represents the case with $|a'_{\alpha\beta}|$ = 0.5 and LIV phase $\phi_{\alpha\beta}$ varied in the range $[-\pi,\pi]$.
We observe that, 

\begin{itemize}
    \item The presence of off-diagonal elements with non-zero phase may pose degeneracy in the measurement of $\delta_{CP}$. The off-diagonal phases appears with $\delta_{CP}$ as seen in equation \ref{Probmue:main}, and they can mimic the effect of CP-violation making the measurement of $\delta_{CP}$ ambiguous.
    
    \item In presence of $a_{e\mu}$ (top-left panel), the CPV sensitivity mostly shows an improvement in the entire range of $\delta_{CP}$. The enhancement is marginally higher for positive values of $\delta_{CP}$. We notice a significant impact of $\phi_{e\mu}$ on the sensitivity via the shaded grey band.
   
    \item In presence of $a_{e\tau}$ (top-right panel) or $a_{\mu\tau}$ (bottom panel), the effect on sensitivity is nominal when $\phi_{e\tau}$ =0. We observe a reasonable impact of non-zero phase $\phi_{e\mu}$ and $\phi_{\mu\tau}$ as represented by the shaded bands.  

\end{itemize}

\subsection{The CP--Precision Sensitivity of DUNE in presence of LIV}\label{sec:CPPrec_sens}
\begin{figure}[!h]
\centering
\begin{minipage}{0.49\textwidth}
  \centering
  \includegraphics[width=1.1\linewidth, height = 5.5cm]{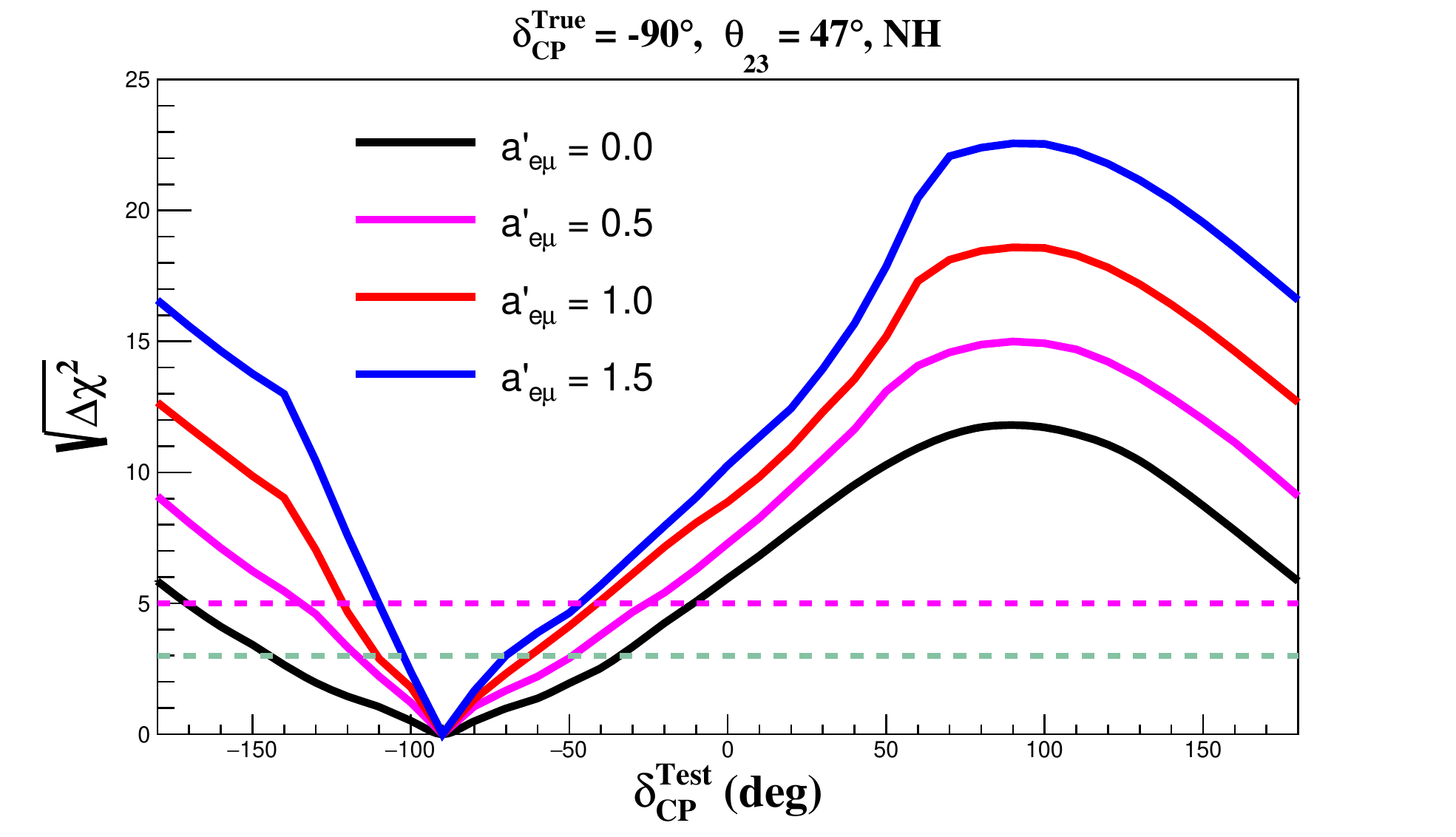}
\end{minipage}
\begin{minipage}{.49\textwidth}
  \centering
  \includegraphics[width=1.1\linewidth, height = 5.5cm]{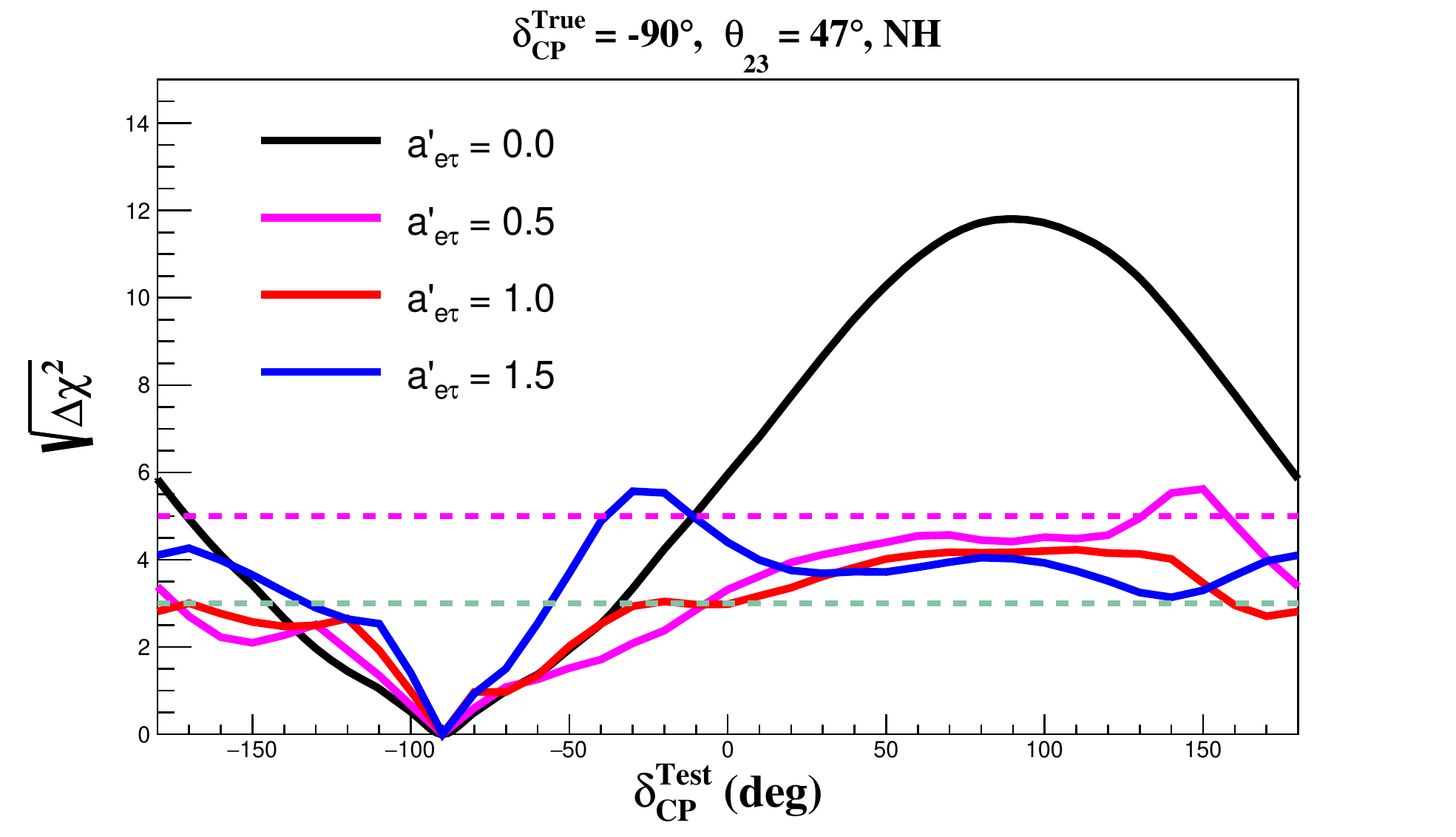}
\end{minipage}
\begin{minipage}{.49\textwidth}
  \centering
  \includegraphics[width=1.1\linewidth, height = 5.5cm]{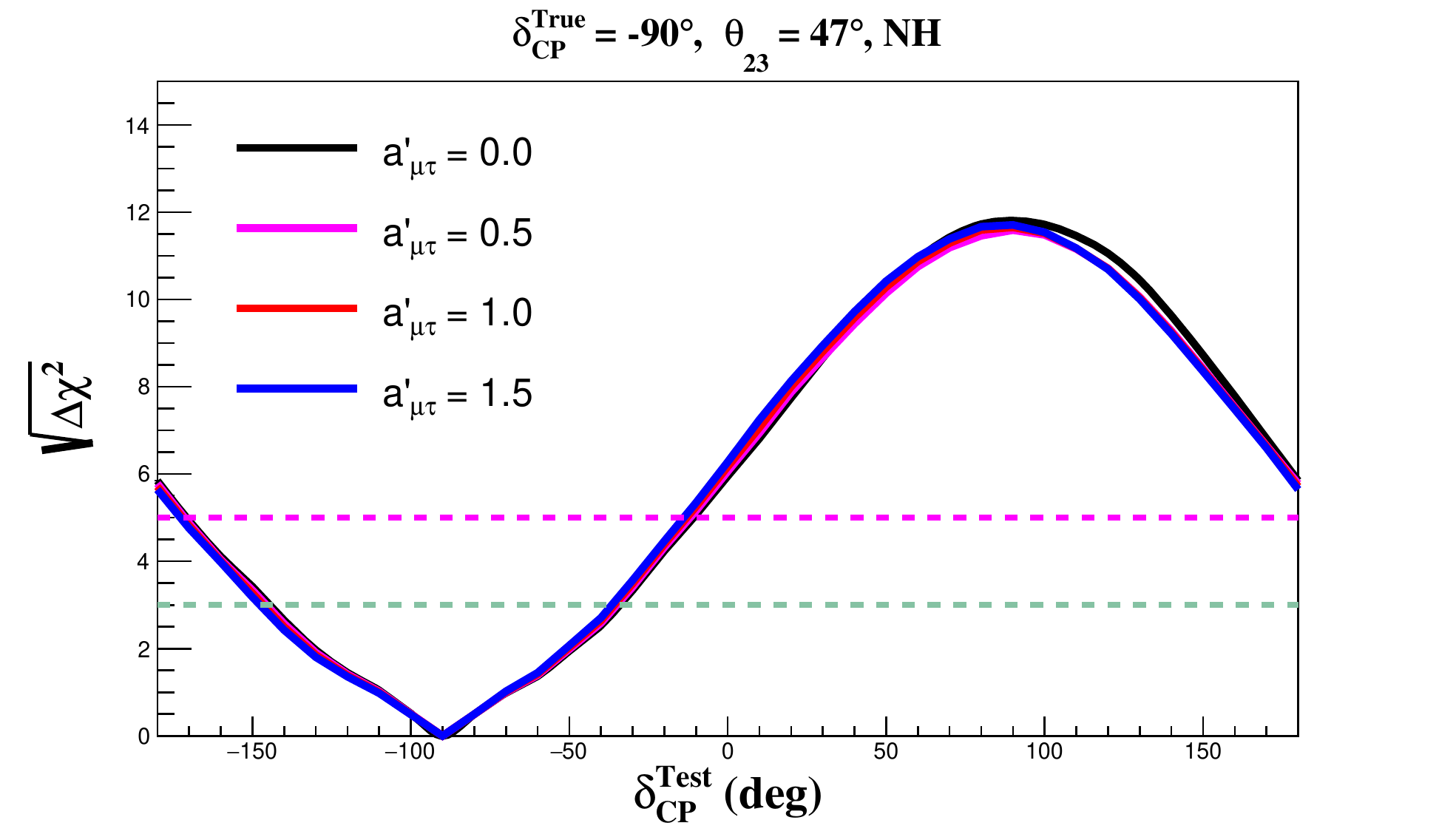}
\end{minipage}
\caption{The CP-precision sensitivity of DUNE in presence of $a_{e\mu}$ (top-left panel), $a_{e\tau}$ (top-right panel) and $a_{\mu\tau}$ (bottom panel). We keep $\phi_{\alpha\beta} = 0^{\circ}$. The parameter $a'_{\alpha\beta}$ represents the values of $a_{\alpha\beta}$ as expressed in equation  \ref{eq:prime_offdiagonal}. The black solid line represents the standard case with no LIV effect. The dashed magenta and green solid line represents the 5$\sigma$ and 3$\sigma$ CL respectively.}
\label{fig:cp_precision}
\end{figure}

We now present the CP-precision sensitivity of DUNE in presence of the LIV parameters $a_{e\mu}$ (top-left panel), $a_{e\tau}$ (top-right panel) and $a_{\mu\tau}$ (bottom panel) in figure \ref{fig:cp_precision}. We look for determining how accurately DUNE can constrain the values of $\delta_{CP}$, considering its true value is known, under the impact of LIV. We have performed a marginalization over the octant $\theta_{23}$, mass squared splitting  $\Delta {m_{31}}^{2}$ and LIV parameter $a'_{\alpha \beta}$ in the range $[0, 2]$. The benchmark values of oscillation parameters used in the analysis are listed in table \ref{tab:parameters}.

The significance $\sqrt{\Delta \chi^{2}}$ is plotted as a function of $\delta_{CP}^{Test}$ in the complete parameter space $[-\pi,\pi]$. The black solid line corresponds to the standard case and the dashed magenta (green) line corresponds to 5$\sigma$ (3$\sigma$) CL. The true value of $\delta_{CP}$ has been fixed at $-\pi/2$. For the standard no-LIV scenario, the CP--precision of DUNE is around $\sim-{90^{\circ}}^{+45^{\circ}}_{-55^{\circ}}$ at 3$\sigma$ CL which is represented by the solid black line. We observe the following from 
figure \ref{fig:cp_precision}.

\begin{itemize}
   
    \item We note a significant enhancement in the CP-precision measurement in presence of $a_{e\mu}$. The enhancement increases with the increase in the strength of $a_{e\mu}$. As example, for $a_{e\mu}$ = 1.5 the CP-precision capability improves as $\sim-{90^{\circ}}^{+42^{\circ}}_{-20^{\circ}}$ at 3$\sigma$ CL.
    
    \item The CP-precision sensitivity mostly deteriorates in the presence of $a_{e\tau}$, particularly for positive $\delta_{CP}$. 
       
    \item The effect of $a_{\mu\tau}$ on the CP-precision measurement capability is marginal.
    \end{itemize}

   \begin{figure}[h]
    \centering
    \begin{minipage}{0.48\textwidth}
      \centering
      \includegraphics[width=1.1\textwidth, height = 5.5 cm]{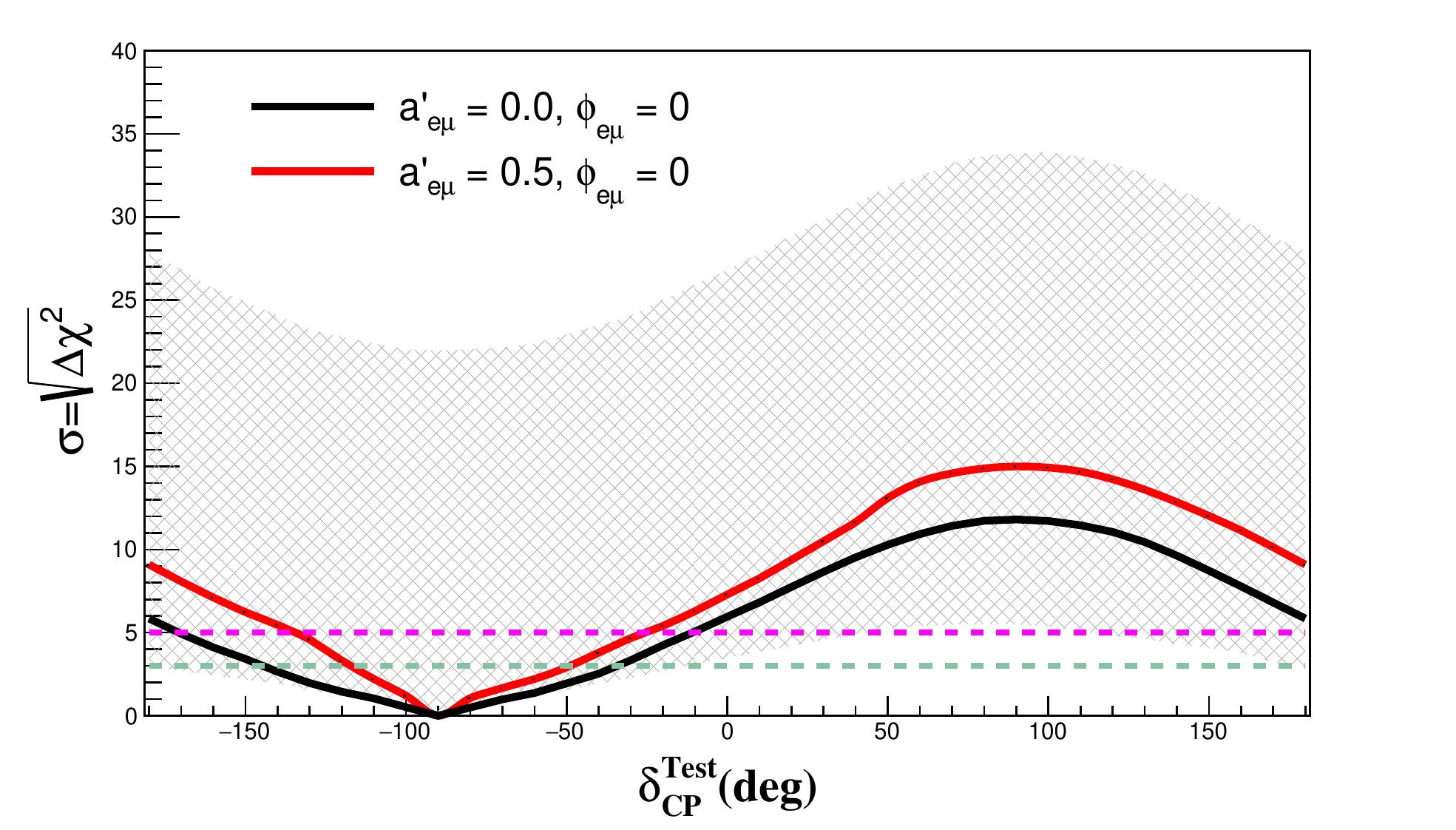}
    \end{minipage}%
        \begin{minipage}{0.48\textwidth}
      \centering
      \includegraphics[width=1.1\textwidth, height = 5.5 cm]{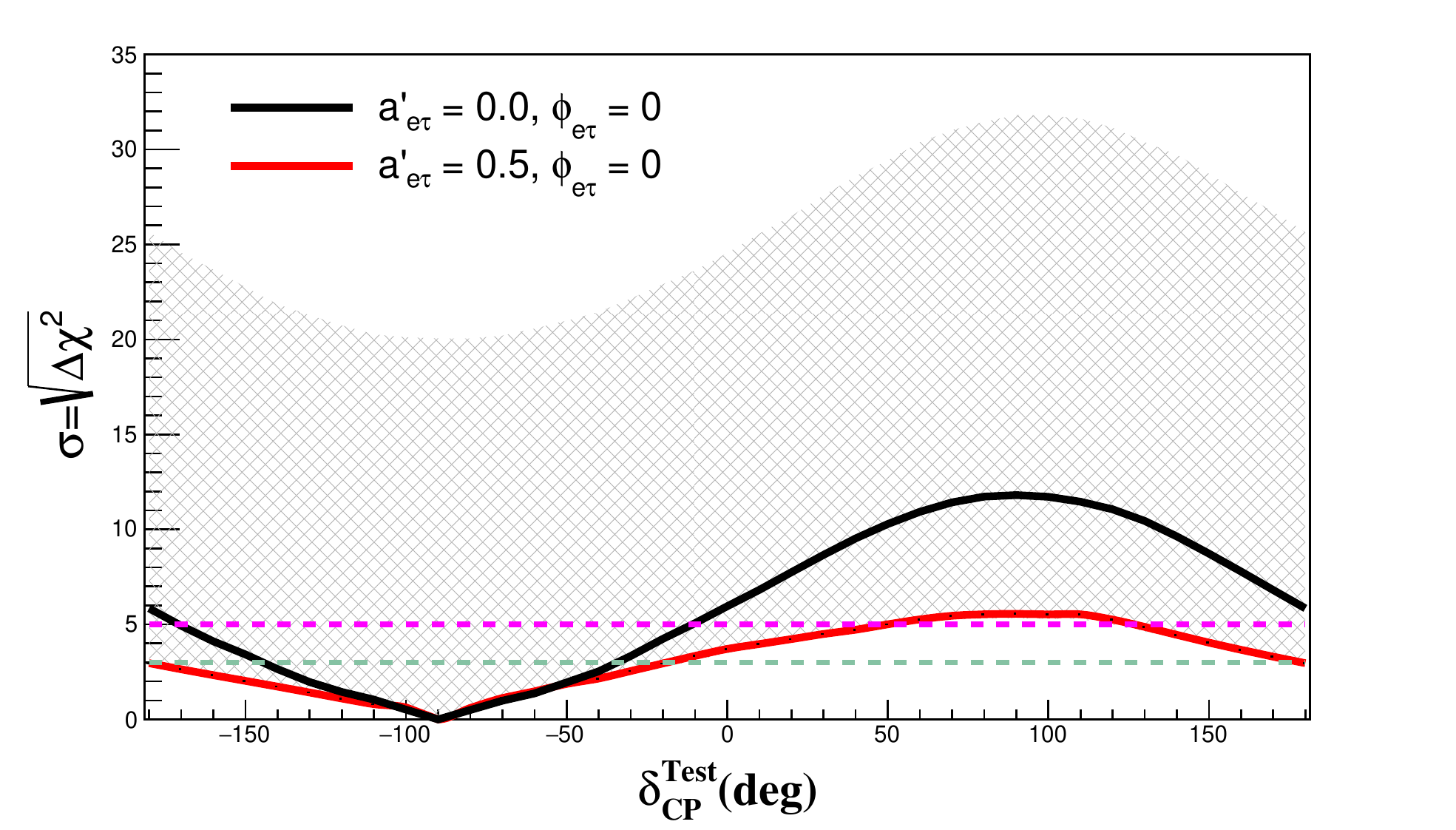}
    \end{minipage}%
    \\
        \begin{minipage}{0.5\textwidth}
      \centering
      \includegraphics[width=1.1\textwidth, height = 5.5 cm]{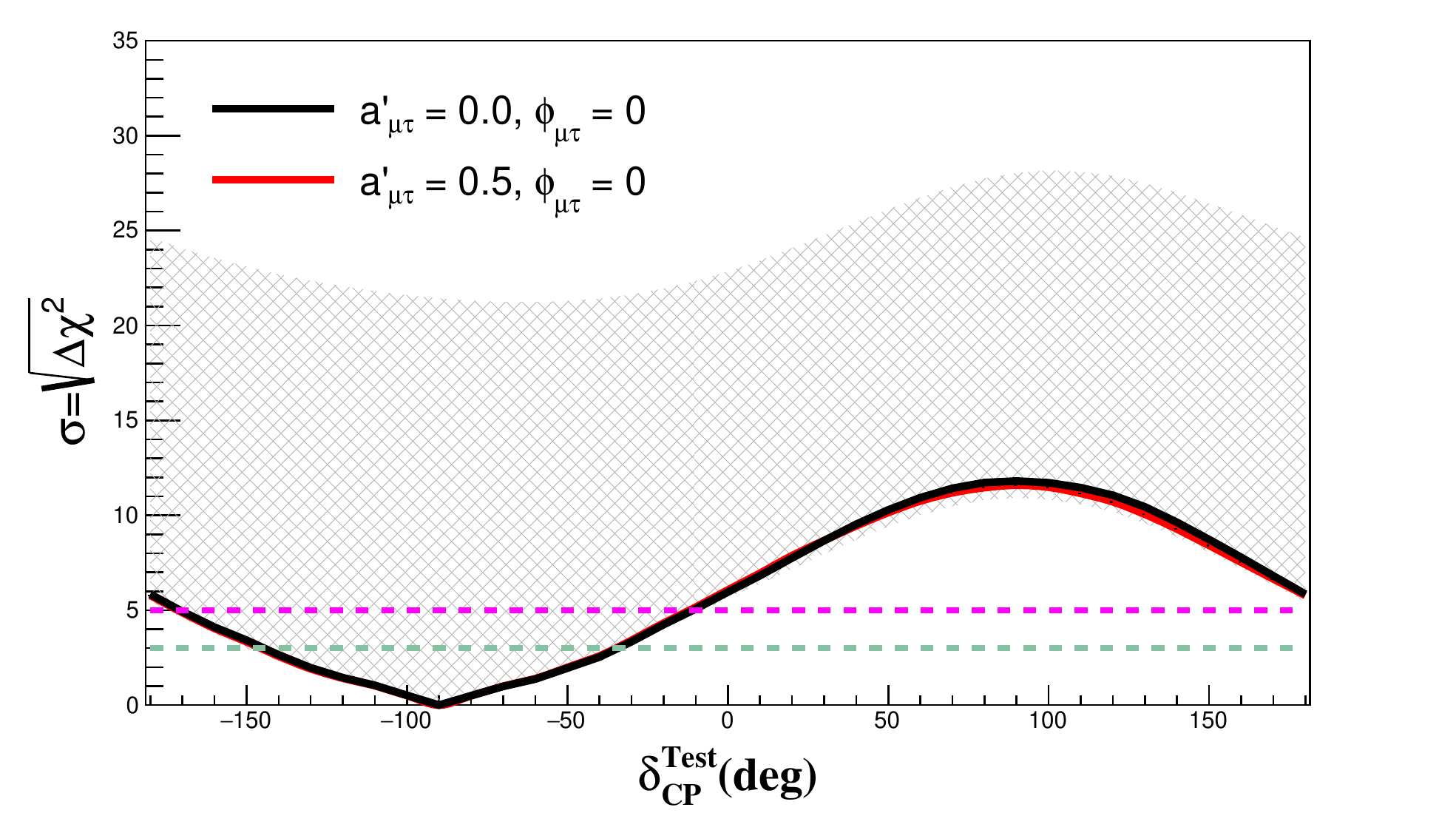}
    \end{minipage}%
    \caption{The CP-precision sensitivity of DUNE in presence of off-diagonal LIV parameters $a_{e\mu}$ (top-left panel), $a_{e\tau}$ (top-right panel) and $a_{\mu\tau}$ (bottom panel). In all sub-figures, the black solid line represents the standard case with no LIV effects whereas the red solid line represents the case with $|a'_{\alpha\beta}| = 0.5$ and $\phi_{\alpha\beta}=0$. The shaded region represents the sensitivities corresponding to varied LIV phases $\phi_{\alpha\beta}$ in the range [$-\pi$,$\pi$] and fixed $|a'_{\alpha\beta}| = 0.5$. The parameter $a'_{\alpha\beta}$ represents the values of $a_{\alpha\beta}$ as expressed in equation \ref{eq:prime_offdiagonal}. The dashed magenta and green line represent $5\sigma$ and $3\sigma$ CL respectively.}
    \label{fig:cpPrec_with_phi}
    \end{figure}

In figure \ref{fig:cpPrec_with_phi}, we show the effects of LIV phases on CP-precision sensitivities. We have additionally marginalized over the parameters $a_{\alpha \beta}$ and $\phi_{\alpha\beta}$ for the calculation of $\Delta \chi^{2}$. The shaded grey band represents the case with $|a'_{\alpha\beta}|$ = 0.5 and the phase $\phi_{\alpha\beta}$ $\in[-\pi,\pi]$. The solid black lines represent the standard case (without--LIV) and the case with $|a'_{\alpha\beta}|$ = 0.5 and $\phi_{\alpha\beta}$ = 0 are shown in solid red line. The magenta and green dashed line represent the $5\sigma$ and $3\sigma$ CL respectively.
We see that, the presence of off-diagonal elements with the LIV phases may induce new degeneracies in the measurement of $\delta_{CP}$ phase. It is due to the occurrence of off-diagonal phases with $\delta_{CP}$ as seen in equation \ref{Probmue:main}. We list our observations from figure \ref{fig:cpPrec_with_phi} as,

\begin{itemize}
    \item In presence of $a_{e\mu}$ (top-left panel), the CP-precision sensitivity gets modified for the complete parameter space of $\delta_{CP}$. The shaded region extended around the zero phase case, indicates a $\phi_{e\mu}$--dependent enhancement/suppression.
    \item The $a_{e\tau}$ parameter (top-right panel) with $\phi_{e\tau}$ = 0 lies in the bottom of the band. The other values of $\phi_{e\tau}$ improves the CP-precision sensitivity.
    \item Similarly, for $a_{\mu\tau}$ (bottom panel), the sensitivity lies at the bottom of the band for $\phi_{\mu\tau}=0$ case. The CP-precision sensitivity gets improved for the non-zero values of the phase $\phi_{\mu\tau}$.  
\end{itemize}
    
\section{Summary and Concluding Remarks}\label{sec:conclusion}

The Lorentz Invariance is a fundamental symmetry of space-time and a violation of this symmetry may be studied as a subdominant effect on the neutrino oscillation probabilities. Currently, we are in the precision era of neutrino physics where precise measurements of mixing parameters are being obtained by various neutrino experiments. Possible non-standard effects like LIV may affect the measurement sensitivity of such oscillation parameters in different neutrino experiments. This also opens up a portal to probe violation of such fundamental symmetries via neutrino oscillations. 

In this work, we have explored the effects of LIV parameters on various measurements at the upcoming DUNE detector. We have performed a $\chi^{2}$ analysis to study the impact of LIV parameters $a_{\alpha\beta}$ on the physics scope of the experiment. We particularly probe the effect on the CP-Violation sensitivity of DUNE in the presence of $a_{e\mu}$, $a_{e\tau}$ and $a_{\mu\tau}$. We see that the presence of $a_{e\mu}$ significantly enhances the CP sensitivity whereas the presence of $a_{e\tau}$ deteriorates the sensitivity. The effects due to $a_{\mu\tau}$ on CP sensitivity is nominal. The presence of a non-zero phase can significantly enhance/suppress the sensitivities. This indicates that the impact of LIV parameters cannot be ignored in long baseline experiments as the sensitivities will get affected in the presence of CPT-odd LIV parameters. We have also performed a CP-precision study where we have explored the impacts of $a_{e\mu}$, $a_{e\tau}$ and $a_{\mu\tau}$, on the $\delta_{CP}$ constraining capability of DUNE. The presence of $a_{e\mu}$ significantly improves the capability to constrain $\delta_{CP}$, whereas $a_{e\tau}$ slightly deteriorates the capability and $a_{\mu\tau}$ shows nominal effects. Also, the addition of off-diagonal phases may induce degeneracies in the measurement of $\delta_{CP}$ phase.

The study of the impact of such sub-dominant effects in neutrino oscillations is very crucial for accurate and precise measurements of neutrino mixing parameters. It is also important to identify the capability of the detectors in observing such non-standard effects if they exist in nature. This in turn will also help us better understand the nature of neutrinos.

\section*{Acknowledgments}
The authors acknowledge the Science and Engineering Research Board (SERB), DST for the grant CRG/2021/002961. AS acknowledges the fellowship received from CSIR-HRDG (09/0796(12409)/2021-EMR-I). AM acknowledges the support of Research and Innovation grant 2021 (DoRD/RIG/10-73/1592-A) funded by Tezpur University. The authors also acknowledge the DST FIST grant SR/FST/PSI-211/2016(C) received by the Department of Physics, Tezpur University.

\section{Appendix}
\subsection{The approximate probability expressions till the second order of matter coefficient a.}\label{App:secondorder}

\begin{itemize}
\item The appearance probability expression in second order in the presence of only $a_{e\mu}$ LIV parameter can be expressed as,
\begin{multline*}
P_{\mu e}(a_{e\mu})=4c_{23}^{2}\left|c_{12}s_{12}\frac{\Delta m_{21}^{2}}{a}+\frac{c_{23}a_{e\mu}}{A_{c}}\right|^{2}sin^{2}(\frac{aL}{4E})\\
+4s_{23}^{2}\left|s_{13}e^{-i\delta}\frac{\Delta m_{31}^{2}}{a}+\frac{s_{23}a_{e\mu}}{A_{c}}\right|^{2}\left(\frac{a}{\Delta m_{31}^{2}-a}\right)^{2}sin^{2}\left(\frac{\Delta m_{31}^{2}-a}{4E}L\right)\\
+8c_{23}s_{23}Re\left[\left(c_{12}s_{12}\frac{\Delta m_{21}^{2}}{a}+\frac{c_{23}a_{e\mu}}{A_{c}}\right)\left(s_{13}e^{i\delta}\frac{\Delta m_{31}^{2}}{a}+\frac{s_{23}a_{e\mu}^{*}}{A_{c}}\right)\right]\\
\times\left(\frac{a}{\Delta m_{31}^{2}-a}\right)sin\left(\frac{aL}{4E}\right)cos\left(\frac{\Delta m_{31}^{2}L}{4E}\right)sin\left(\frac{\Delta m_{31}^{2}-a}{4E}L\right)\\
+8c_{23}s_{23}Im\left[\left(c_{12}s_{12}\frac{\Delta m_{21}^{2}}{a}+\frac{c_{23}a_{e\mu}}{A_{c}}\right)\left(s_{13}e^{i\delta}\frac{\Delta m_{31}^{2}}{a}+\frac{s_{23}a_{e\mu}^{*}}{A_{c}}\right)\right]\\
\times\left(\frac{a}{\Delta m_{31}^{2}-a}\right)sin\left(\frac{aL}{4E}\right)sin\left(\frac{\Delta m_{31}^{2}L}{4E}\right)sin\left(\frac{\Delta m_{31}^{2}-a}{4E}L\right).
\end{multline*}
\begin{equation}\label{eq:2ndOrder_Pmue}
\end{equation}

\item The appearance probability expression in second order in the presence of only $a_{e\tau}$ LIV parameter can be expressed as,
\begin{multline*}
P_{\mu e}(a_{e\tau})=4c_{23}^{2}\left|c_{12}s_{12}\frac{\Delta m_{21}^{2}}{a}-\frac{s_{23}a_{e\tau}}{A_{c}}\right|^{2}sin^{2}(\frac{aL}{4E})\\
+4s_{23}^{2}\left|s_{13}e^{-i\delta}\frac{\Delta m_{31}^{2}}{a}+\frac{c_{23}a_{e\tau}}{A_{c}}\right|^{2}\left(\frac{a}{\Delta m_{31}^{2}-a}\right)^{2}sin^{2}\left(\frac{\Delta m_{31}^{2}-a}{4E}L\right)\\
+8c_{23}s_{23}Re\left[\left(c_{12}s_{12}\frac{\Delta m_{21}^{2}}{a}-\frac{s_{23}a_{e\tau}}{A_{c}}\right)\left(s_{13}e^{i\delta}\frac{\Delta m_{31}^{2}}{a}+\frac{c_{23}a_{e\tau}^{*}}{A_{c}}\right)\right]\\
\times\left(\frac{a}{\Delta m_{31}^{2}-a}\right)sin\left(\frac{aL}{4E}\right)cos\left(\frac{\Delta m_{31}^{2}L}{4E}\right)sin\left(\frac{\Delta m_{31}^{2}-a}{4E}L\right)\\
+8c_{23}s_{23}Im\left[\left(c_{12}s_{12}\frac{\Delta m_{21}^{2}}{a}-\frac{s_{23}a_{e\tau}}{A_{c}}\right)\left(s_{13}e^{i\delta}\frac{\Delta m_{31}^{2}}{a}+\frac{c_{23}a_{e\tau}^{*}}{A_{c}}\right)\right]\\
\times\left(\frac{a}{\Delta m_{31}^{2}-a}\right)sin\left(\frac{aL}{4E}\right)sin\left(\frac{\Delta m_{31}^{2}L}{4E}\right)sin\left(\frac{\Delta m_{31}^{2}-a}{4E}L\right).
\end{multline*}
\begin{equation}\label{eq:2ndOrder_etau}
\end{equation}

\item The disappearance probability expression in second order in the presence of only $a_{\mu\tau}$ LIV parameter can be expressed as,
\begin{multline*}
P_{\mu\mu}(a_{\mu\tau})=1-4c_{23}^{2}s_{23}^{2}sin^{2}\left(\frac{\Delta m_{31}^{2}L}{4E}\right)\\
+2c_{23}^{2}s_{23}^{2}\left\{ c_{12}^{2}A_{c}\frac{\Delta m_{21}^{2}}{a}-2c_{23}s_{23}(a_{\mu\tau}+a_{\mu\tau}^{*})+s_{13}^{2}A_{c}\frac{\Delta m_{21}^{2}}{a}\right\} \frac{aL}{2EA_{c}}sin\left(\frac{\Delta m_{31}^{2}L}{2E}\right)\\
-8c_{23}s_{23}\left(c_{23}^{2}-s_{23}^{2}\right)Re\left[c_{23}^{2}a_{\mu\tau}-s_{23}^{2}a_{\mu\tau}^{*}-c_{12}s_{12}s_{13}A_{c}e^{-i\delta}\frac{\Delta m_{21}^{2}}{a}\right]\frac{a}{\Delta m_{31}^{2}A_{c}}sin^{2}\left(\frac{\Delta m_{31}^{2}L}{4E}\right)\\
-\frac{c_{23}^{2}s_{23}^{2}}{A_{c}}\left\{ c_{12}^{2}A_{c}\frac{\Delta m_{21}^{2}}{a}-2c_{23}s_{23}\left(a_{\mu\tau}+a_{\mu\tau}^{*}\right)\right\}\} ^{2}\left(\frac{aL}{2E}\right)^{2}cos\left(\frac{\Delta m_{31}^{2}L}{2E}\right)\\
+\frac{4c_{23}s_{23}}{A_{c}}\left(c_{23}^{2}-s_{23}^{2}\right)\left\{ c_{12}^{2}A_{c}\frac{\Delta m_{21}^{2}}{a}-2c_{23}s_{23}\left(a_{\mu\tau}+a_{\mu\tau}^{*}\right)\right\}\\
\times Re\left[\frac{c_{23}^{2}a_{\mu\tau}-s_{23}^{2}a_{\mu\tau}^{*}}{A_{c}} \right]\left[\frac{a}{\Delta m_{31}^{2}}\frac{aL}{2E}sin\left(\frac{\Delta m_{31}^{2}L}{2E}\right)-2\left(\frac{a}{\Delta m_{31}^{2}}\right)^{2}sin^{2}\left(\frac{\Delta m_{31}^{2}L}{4E}\right)\right]\\
-4c_{23}^{2}s_{23}^{2}\left|c_{23}^{2}a_{\mu\tau}-s_{23}^{2}a_{\mu\tau}^{*}\right|^{2}\frac{a}{\Delta m_{31}^{2}A_{c}^{2}}\frac{aL}{2E}sin\left(\frac{\Delta m_{31}^{2}L}{2E}\right)\\
-4\left(c_{23}^{2}-s_{23}^{2}\right)^{2}\left|c_{23}^{2}a_{\mu\tau}-s_{23}^{2}a_{\mu\tau}^{*}\right|^{2}\left(\frac{a}{\Delta m_{31}^{2}A_{c}}\right)^{2}sin^{2}\left(\frac{\Delta m_{31}^{2}L}{4E}\right)\\
+16c_{23}^{2}s_{23}^{2}\left(Re\left[c_{23}^{2}a_{\mu\tau}-s_{23}^{2}a_{\mu\tau}^{*}\right]\right)^{2}\left(\frac{a}{\Delta m_{31}^{2}A_{c}}\right)^{2}sin^{2}\left(\frac{\Delta m_{31}^{2}L}{4E}\right).
\end{multline*}

\begin{equation}\label{eq:2ndOrder_Pmumu}
\end{equation}

\end{itemize}

\newpage

\bibliographystyle{JHEP}
\bibliography{LIV}
\end{document}